\documentstyle[12pt,leqno]{article}
\font\tenttt=cmtt10 scaled \magstep2
\font\seventtt=cmtt8 scaled \magstep1
\font\fivettt=cmtt8
\newfam\tttfam \def\ttt{\fam\tttfam\tenttt}
\textfont\tttfam=\tenttt
\scriptfont\tttfam=\seventtt
\scriptscriptfont\tttfam=\fivettt

\newtheorem{theorem}{Theorem}[section]
\newtheorem{definition}[theorem]{Definition}
\newtheorem{corollary}[theorem]{Corollary}

\newtheorem{odstavec}[theorem]{}
\newtheorem{example}[theorem]{Example}
\newtheorem{lemma}[theorem]{Lemma}

\newtheorem{proposition}[theorem]{Proposition}

\catcode`\@=11
\def\@begintheorem#1#2{\it \trivlist \item[\hskip
 \labelsep{\bf #1\ #2.}]}
\def\@opargbegintheorem#1#2#3{\it \trivlist\item[\hskip%
 \labelsep{\bf #1\ #2.\ (#3)}]}
\def\@endtheorem{\endtrivlist}

\def\@listI{\leftmargin\leftmargini \parsep 1pt plus 2.5pt
 minus 1pt\topsep 5pt plus 2pt minus 3pt%
 \itemsep 0pt plus 2.5pt minus 1pt}
\let\@listi\@listI
\@listi

\def\@sect#1#2#3#4#5#6[#7]#8{\ifnum #2>\c@secnumdepth%
 \def \@svsec {}\else \refstepcounter {#1}\edef \@svsec%
 {\csname the#1\endcsname. \hskip .1em }\fi \@tempskipa%
 #5\relax \ifdim \@tempskipa >\z@ \begingroup #6\relax%
 \@hangfrom {\hskip #3\relax \@svsec }{\interlinepenalty%
 \@M #8\par }\endgroup \csname #1mark\endcsname {#7}%
 \addcontentsline {toc}{#1}{\ifnum #2>\c@secnumdepth%
 \else \protect \numberline {\csname the#1\endcsname. }%
 \fi #7}\else \def \@svsechd {#6\hskip #3\@svsec #8%
 \csname #1mark\endcsname {#7}\addcontentsline {toc}{#1}%
 {\ifnum #2>\c@secnumdepth \else \protect \numberline%
 {\csname the#1\endcsname. }\fi #7}}\fi \@xsect {#5}}

\def\section{\@startsection {section}{1}{\z@ }%
 {-3.5ex plus -1ex minus -.2ex}{2.3ex plus .2ex}{\sc }}

\def\thebibliography#1{%
 \section *{References\@mkboth {REFERENCES}{REFERENCES}}%
 \list {[\arabic {enumi}]}{\settowidth \labelwidth {[#1]}%
 \leftmargin \labelwidth \advance \leftmargin \labelsep %
 \usecounter {enumi}} \def \newblock %
 {\hskip .11em plus .33em minus -.07em} \sloppy \clubpenalty 4000%
 \widowpenalty 4000 \sfcode`\.=1000\relax}

\def\@maketitle{%
 \newpage \null \vskip 2em
 \begin{center}{\Large\sc \@title \par }
 \vskip 1.5em
 {\large \lineskip .5em
 \begin {tabular}[t]{c}\@author
 \end{tabular}\par }
 \end{center}\par \vskip 1.5em}

\def\abstract{%
\if@twocolumn \section *{Abstract}
 \else \small\quotation\noindent{\bf Abstract.}\fi}

\def\ps@myheadings{\let\@mkboth\@gobbletwo
\def\@oddhead{\ifnum\thepage=0 \hfill \else
\rm [September 30, 1994]\hfil\thepage\fi}
\def\@oddfoot{\hfil}
\def\@evenhead{\rm \thepage\hfil\sl\leftmark\hbox {}}%
\def\@evenfoot{}\def\sectionmark##1{}\def\subsectionmark##1{}}
\catcode`\@=13

\def\qed{\hspace*{\fill}
\mbox{\hphantom{mm}\rule{0.25cm}{0.25cm}}\\
\par\vskip-5mm}

\def\eodst{\hspace*{\fill}
\mbox{\hphantom{mm}{\bf --}}\\
\par\vskip-14mm\hphantom{m}}

\def\bbbR{{\mbox{R \hskip-5mm R}}}
\def\tr{{\cal T}}
\def\vert{{\bf v}}
\def\fr{{\cal F}}
\def\sp{\mbox{\rm Span}}
\def\S{{\cal S}}
\def\trb{\tr^{2,3_1}}
\def\ot{\otimes}
\def\mod{\mbox{ mod }}
\def\modn#1{\mbox{ \rm mod}_{#1}\ (S,T)}
\def\op{\oplus}
\def\trwb{\tr^{wb}} \def\P{{\cal P}}
\def\A{{\cal A}} \def\B{{\cal B}} \def\C{{\cal C}}
\def\tT{{\ttt T}} \def\tS{{\ttt S}} \def\tR{{\ttt R}}
\def\sfT{{\sf T}} \def\sfS{{\sf S}}
\def\id{1\!\!1}
\def\out{\mbox{\rm out}}
\def\inp{\mbox{\rm inp}}
\def\c{\circ}
\def\aa#1#2{{\alpha}_{{#1},{#2}}}
\def\b{\bullet}
\def\E{{\bf E}}
\def\vect{{\bf Vect}}
\def\coll{{\bf Coll}}
\def\oper{{\bf Oper}}
\def\bbox{\raisebox{-.5mm}{{\mbox{\large $\Box$}}}}
\def\val{{\mbox{\rm val}}}
\def\prez#1#2{\langle #1;#2\rangle}
\def\mut{\mu^{\sfT}}
\def\etat{\eta^{\sfT}}
\def\mus{\mu^{\sfS}}
\def\etas{\eta^{\sfS}}
\def\cd{\odot}
\def\bk{{\bf k}}
\def\whk{{\widehat{\bk \oplus \bk}}}
\def\ii{(\infty,\infty)}
\def\whkmod{\whk\mbox{ mod }\bk_0}

\begin{document}
\baselineskip18pt
\bibliographystyle{plain}

\title{Distributive laws and the Koszulness}
\author{Martin Markl}

\maketitle

\noindent
{\sc Introduction}

\vskip2mm
The basic motivation for our work was the following result
of Getzler and Jones~\cite{getzler-jones:preprint}.
Let $\C_n = \{\C_n(m);\ m\geq 1\}$ be the famous little $n$-cubes
operad of Boardman and Vogt~\cite[Definition~4.1]{may:1972} and let
${\bf e}_n =\{{\bf e}_n(m);\ m\geq 1\}$ be its homology operad,
${\bf e}_n(m):= H(\C_n(m))$. Then the operad ${\bf e}_n$ is
Koszul in the sense of~\cite{ginzburg-kapranov:preprint}.

Several comments are in order. The Koszulness of operads
is a certain homological property which is
an analog of the similar classical property of
associative algebras. Koszul operads share many nice properties, for
example, there exist an explicit and effective way to compute the
(co)homology of algebras over these operads,
see~\cite{ginzburg-kapranov:preprint,zebrulka}.

Both the little cubes operad and the operad ${\bf e}_n$ are intimately
related to configuration spaces, namely, ${\bf e}_n(m) =
H({\bf F}_n(m))$, where ${\bf F}_n(m)$ denotes the space of
configurations of $m$ distinct points in ${\bbbR}^n$. The physical
relevance of the operad ${\bf e}_n$ is given by the fact that
some spaces playing an important r\^ole in closed string field theory
are compactifications (of various types) of the
configuration space ${\bf F}_n$, see~\cite{KSV}.

In their original proof of the above mentioned statement, Getzler and
Jones used the Fulton-MacPherson
compactification~\cite{fulton-macpherson:anm94}
${\sf F}_n$ of ${\bf F}_n$. Each ${\sf F}_n(m)$ is a real smooth
manifold with corners and the basic trick in their proof was to
replace, using a spectral sequence associated with the stratification
of ${\sf F}_n(m)$,
the homological definition of the Koszulness by a purely combinatorial
property of the structure of the strata of ${\sf F}_n(m)$.

The operad ${\bf e}_n$
describes so-called $n$-algebras (in the
terminology of~\cite{getzler-jones:preprint}) which are, roughly
speaking, Poisson algebras where the Lie bracket is of
degree $n-1$, especially, $n$-algebras are algebras with a
{\em distributive law\/}. As we already know from our previous
work with T.~Fox~\cite{fox-markl:preprint}, a
distributive law induces a spectral sequence for the related
cohomology. This observation stimulated us to look for an
alternative, purely algebraic proof of the above mentioned
theorem of Getzler and Jones. As usual, we then
found that there are many interesting ramifications on the way.

Distributive laws were introduced and studied,
in terms of triples, by J.~Beck in~\cite{beck:LNM80}.
They provide a way of composing two algebraic structures
into a more complex one. For example, a Poisson algebra
structure on a vector space $V$ consist of a Lie algebra bracket
$[-,-]$ (denoted sometimes more traditionally as $\{-,-\}$) and of an
associative commutative multiplication
$\hskip.5mm\cdot\hskip.5mm$. These two operations are related by a
`distributive law' $[a\cdot b,c]= a\cdot [b,c]+ [a,c]\cdot b$.

Our first aim will be to understand distributive laws in terms of
operads. A distributive law for an operad will be
given by a certain map which has to satisfy a very explicit and
verifiable `coherence condition', see Definition~\ref{16}. We then
prove that an algebra over an operad with a distributive law is and
algebra with a distributive law in the sense of J.~Beck
(Theorem~\ref{48}).
This enables one to
construct new examples of algebras with a distributive law, see
Examples~\ref{49} and~\ref{47}.

Our next aim will be to prove that an operad $\C$ constructed from
operads $\A$ and $\B$ via a distributive law is Koszul if $\A$ and
$\B$ are (Theorem~\ref{44}).
As an immediate corollary we get the above mentioned result
of Getzler and Jones (Corollary~\ref{50}).

Statements about operads are usually motivated by the
corresponding statements
about associative algebras.
This paper gives an example of the inverse phenomena: distributive law
for operads (motivated by the Beck's definition for triples) leads us
to the definition of a distributive law for associative algebras as
of a process which ties together two associative algebras
into a third one (Definition~\ref{15}).
This gives a method to construct new examples of Koszul
algebras.

It has been observed at many places that operads behave similarly as
associative algebras and that there is a dictionary for definitions
and statements about these objects. We believe that this is a
consequence of the fact that both objects are associative algebras in
a certain monoidal category -- associative algebras are (well)
associative algebras in the monoidal category of vector spaces with
the monoidal structure given by the tensor product $\otimes$
while operads are associative algebras in the monoidal
category of collections with the monoidal structure given by the
operation $\cd$, see Proposition~\ref{27}. On the other hand, we do
not know any formal `machine' translating definitions/theorems
for associative algebras to definitions/theorems for operads and vice
versa; the problem seems to be related with the fact that the
bifunctor $\cd$ is not linear in the second variable.

In the present paper which treats simultaneously both the associative
algebra and the operad cases we formulate our
definitions and theorems for
associative algebras first and then for operads. We hope this scheme
will help the reader to understand better the statements for operads,
the associative algebra case is usually easier. We also hope that
the reader will enjoy the following table showing the
correspondence of definitions and statements.

\vskip4mm
$\mbox{Definition of a distributive law}
\left\{
\begin{array}{l}
\mbox{for associative algebras: Definition~\ref{15}}\\
\mbox{for operads: Definition~\ref{16}}
\end{array}
\right.$%\}

$\mbox{Coherence property for a distributive law}
\left\{
\begin{array}{l}
\mbox{for associative algebras: Theorem~\ref{8}}\\
\mbox{for operads: Theorem~\ref{13}}
\end{array}
\right.$%\}

$\mbox{Definition of the Koszulness}
\left\{
\begin{array}{l}
\mbox{for associative algebras: paragraph~\ref{42}}\\
\mbox{for operads: paragraph~\ref{43}}
\end{array}
\right.$%\}

$\mbox{Koszulness versus a distributive law}
\left\{
\begin{array}{l}
\mbox{for associative algebras: Theorem~\ref{21}}\\
\mbox{for operads: Theorem~\ref{44}}
\end{array}
\right.$%\}

\vskip7mm
\noindent
Plan of the paper:

\noindent
\hphantom{mmmm}1. Basic notions\hfill\break\noindent
\hphantom{mmmm}2. Distributive laws\hfill\break\noindent
\hphantom{mmmm}3. Distributive laws and triples\hfill\break\noindent
\hphantom{mmmm}4. Explicit computations and
examples\hfill\break\noindent
\hphantom{mmmm}5. Distributive laws and the Koszulness

\section{Basic notions}

We will keep the following convention throughout the paper.
Capital roman letters ($A$,
$B$, ...) will denote associative algebras, calligraphic letters
($\A$, $\B$, ...) will denote operads, `typewriter' capitals ($\tT$,
$\tS$, ...) will denote trees and, finally, `sans serif' capitals
($\sfT$, $\sfS$, ...) will denote triples.
All algebraic objects are assumed to be defined over a
fixed field $\bk$ which
is, to make the life easier, supposed to be of characteristic zero.

We hope that the notion of an operad and of an algebra over an operad
already became a part of a common knowledge, we thus only briefly
introduce the necessary notation.
By an {\em nonsymmetric operad\/} we mean a nonsymmetric
operad in the monoidal
category $\vect$ of graded vector spaces, i.e.~a
sequence $\S = \{\S(n); n\geq 1\}$ of graded vector spaces
together with degree zero linear maps
\[
\gamma=\gamma_{m_1,\ldots,m_l}:\S(l)\ot\S(m_1)\ot\cdots\ot\S(m_l)
\longrightarrow
\S(m_1+\cdots+m_l),
\]
given for any $l,m_1,\ldots,m_l \geq 1$,
satisfying the usual axioms~\cite[Definition~3.12]{may:1972}. We
also suppose the existence of the unit $1\in \S(1)$ with the property
that
$\gamma(\mu;1,\ldots,1)= \mu$ for each $\mu \in \S(m)$, $m\geq 1$.
Similarly, a {\em symmetric operad\/} will be an operad in the
symmetric monoidal category $\vect$ of graded vector spaces,
i.e.~a structure consisting of the above data plus an action
of the symmetric group $\Sigma_n$ on $\S(n)$ given for any $n\geq 2$,
which has again to satisfy the usual
axioms~\cite[Definition~1.1]{may:1972}. We always
assume that $\S(1)=\bk$ and that algebra structure on $\bk$
coincides, under this identification, with the algebra structure on
$\S(1)$ induced from the operad structure of $\S$.

We will try to discuss both symmetric and nonsymmetric cases
simultaneously whenever possible. We also will
not mention explicitly the grading given
by the grading of underlying vector spaces if not necessary.
For an operad $\S$ we often use also the `nonunital' notation
based on the composition maps $-\c_i- : \S(m)\ot \S(n)\to \S(m+n-1)$,
given, for any $m,n\geq 1$, $1\leq i\leq m$, by
\[
\mu\c_i\nu:=\gamma(\mu;1,\ldots,1,\nu,1,\ldots,1)
\mbox{ ($\nu$ at the $i$-th place)}.
\]
These maps satisfy, for each $f\in \S(a)$, $g\in \S(b)$ and
$h\in \S(c)$, the relations~\cite{zebrulka}
\begin{equation}
\label{25}
(f\c_j g)\c_i h =
\left\{
\begin{array}{ll}
(-1)^{|h|\cdot|g|}\cdot (f\c_i h)\c_{j+c-1}g,&
1\leq i\leq j-1,
\\
f\c_j(g\c_{i-j+1}h), &
j\leq i \leq b+j-1, \mbox{ and}
\\
(-1)^{|h|\cdot|g|}\cdot (f\c_{i-b+1}h)\c_j g,&
i\geq j+b.
\end{array}
\right.%}
\end{equation}

By a {\em collection\/} we mean~\cite{ginzburg-kapranov:preprint}
a sequence $\{E(n);\
n\geq 2\}$ of graded vector spaces; in the symmetric case we suppose
moreover that each $E(n)$ is equipped with an action of the symmetric
group $\Sigma_n$. Let $\coll$ denote the category of collections an
let $\oper$ denote the category of operads. We have the `forgetful'
functor $\bbox : \oper \to \coll$ given by $\bbox(\S)(m):= \S(m)$ for
$m\geq 2$. The functor $\bbox$ has a left
adjoint $\fr:\coll \to \oper$ and the operad $\fr(E)$ is
called the {\em free operad\/} on the collection
$E$~\cite{ginzburg-kapranov:preprint}, see
also~\ref{29}. A very explicit description of $\fr(E)$ using trees is
given in~\ref{29}.
Let us state without proofs some elementary properties of free
operads.

\begin{odstavec}{\rm{\bf --}
\label{26}
Suppose the collection $E$ decomposes as $E=\bigoplus_{i=1}^NE_i$,
meaning, of course, that $E(m)= \bigoplus_{i=1}^NE_i(m)$ for each
$m\geq 2$, the decomposition being $\Sigma_m$-invariant in the
symmetric case. Then
$\fr(E)$ is naturally $N$-multigraded,
$\fr(E)=\bigoplus_{i_1,\ldots,i_N}\fr_{i_1,\ldots,i_N}(E)$, with the
multigrading characterized by the following two properties.
\begin{enumerate}
\item[(i)]
$\fr_{0,\ldots,0}(E)= \fr(E)(1)= \bk$ and
$E_i= \fr_{0,\ldots,0,1,0,\ldots,0}(E)$ (1 at the $i$-th
place), $1\leq i\leq N$.
\item[(ii)]
Let $m,n\geq 1$, $1\leq l \leq m$ and let
$a \in \fr_{i_1,\ldots,i_N}(E)(m)$,
$b\in \fr_{j_1,\ldots,j_N}(E)(n)$. Then $a \c_l b
\in \fr_{k_1,\ldots,k_N}(E)(m+n-1)$ with $k_i = i_i +j_i$ for
$1\leq i\leq N$.
\end{enumerate}
Especially, the trivial decomposition of $E$ gives the grading
$\fr(E)=\bigoplus_{i\geq 0} \fr_i(E)$.
\eodst}\end{odstavec}

Let $E = U\op V$. Denote by $U\cd V$ the subcollection of $\fr(E)$
generated by (= the smallest subcollection containing)
elements of the form $\gamma(u,v_1,\ldots,v_m)$, $u\in U(m)$
and $v_i\in V(n_i)$, $1\leq i\leq m$.
More explicit description of $U\cd V$ can be found in the
proof of Proposition~\ref{38}. The following statement was
formulated for example in~\cite{smirnov:USSRSb82}.

\begin{proposition}
\label{27}
The operation $\cd$ introduced above defines on $\coll$ a structure of
a strict monoidal category. An operad is then an
associative unital algebra in this category.
\end{proposition}

The above should be compared with the properties of the
{\em free associative algebra\/} $F(X)$ on a $K$-vector space $X$. It
 can be constructed by taking $F(X) :=
\bigoplus_{n\geq 0}T^n(X)$, where $T^n(-)$ denotes the $n$-th
tensor power of $X$ over $K$ with the convention that
$T^0(X):=\bk$. The multiplication is defined in an obvious way.
The analog of the multigrading of~\ref{26} is clear. If $X =
\bigoplus_{i=1}^nX_i$ then $F_{i_1,\ldots,i_N}(X)$ is generated by
monomials $x_1\ot \cdots \ot x_{i_1+\ldots+i_N}$ such that $x_j
\in X_k$ for exactly $i_k$ indices $j$, $1\leq k\leq N$.
The grading given by the
trivial decomposition coincides with the usual one, $F_n(X)=
T^n(X)$. The operation $\cd$ of Proposition~\ref{27} corresponds to
the usual tensor product.

\begin{odstavec}{\rm {\bf -- }
\label{36}
By a {\em presentation\/} of an associative
algebra $A$ we mean a vector space $X$ and a subspace $R\subset F(X)$
such that $A= F(X)/(R)$, where $(R)$ denotes the ideal
generated by $R$ in $F(X)$. In this situation we write $A=
\langle X; R\rangle$.
We say that $A$ is {\em quadratic\/} if it has a presentation
$\langle X ; R\rangle$ with $R \subset F_2(X)$. Because of the
homogeneity of the relations, a quadratic algebra
$A$ is naturally graded, $A = \bigoplus_{i\geq 0}A_i$, the grading
being
induced by that of $F(X)$.
There is a very explicit description of $A_n$ in terms of $X$ and
$R$. Let $\tilde R$ denotes another, `abstract' copy of
$R$ and let $\iota : X\op \tilde R\to F(X)$ be the obvious
map of vector spaces. Then $\iota$ induces, by the universal
property, an algebra homomorphism $h :F(X\op \tilde R)\to F(X)$ and
we have $A_n = F_n(X)/h(F_{n-2,1}(X\op \tilde R))$. Still more
explicitly,
$h(F_{n-2,1}(X\op \tilde R)) = \sp(R_{s,n};\ 0\leq s\leq n-2)$,
where $R_{s,n} := T^s(X)\otimes R
\otimes T^{n-s-2}(X) \subset T^n(X)=F_n(X)$.
\eodst}\end{odstavec}

\begin{odstavec}{\rm {\bf --}
\label{28}
By a {\em presentation\/} of an operad $\S$ we mean a
collection $E$ and a subcollection $R\subset \fr(E)$
such that $\S= \fr(E)/(R)$, where $(R)$ denotes the ideal
generated by $R$ in $\fr(X)$.
We write $\S = \langle E;R\rangle$. An operad $\S$ is
{\em quadratic\/} if there exists a collection $E$ with $E(n)=0$ for
$n\not= 2$, and a subcollection $R \subset \fr(E)(3)$
such that $\S = \langle E;R\rangle$. Similarly as quadratic
associative algebras, quadratic operads are graded, $\S =
\bigoplus_{i\geq 0}\S_i$. We have, moreover, a similar description of
the pieces $\S_n$ as in the associative algebra case. Namely, let
$\tilde R$ be the identical copy of the collection $R$ and let
$h:\fr(E\op \tilde R)\to \fr(E)$ be the map induced by the obvious
map $\iota : E\op \tilde R\to \fr(E)$ of collections.
Then $\S_n = \fr_n(E)/h(\fr_{n-2,1}(E\op \tilde R))$.
For a more explicit description, see~\ref{37}.
\eodst}\end{odstavec}

\begin{odstavec}{\rm {\bf --}
\label{29}
There exists a useful way to describe free operads using
trees~\cite{ginzburg-kapranov:preprint,getzler-jones:preprint}.
In the {\em nonsymmetric case} we shall use the set $\tr$ of
planar trees. By $\tr_n$ we denote the subset of $\tr$ consisting of
trees having $n$ input edges. Let
$\vert(\tT)$ denote the set of vertices of a tree $\tT \in \tr$ and
let, for $v\in \vert(\tT)$, $\val(v)$ denote the number of input
edges of $v$. For a collection $E = \{E(n);n\geq 2\}$ we put
\[
E(\tT):=
\bigotimes_{v\in\vert(\tT)}E(\val(v)).
\]
We may interpret the elements of $E(\tT)$
as `multilinear' colorings of the vertices of $\tT$ by the elements of
$E$. The free operad $\fr(E)$ on $E$ may be then defined as
\begin{equation}
\label{30}
\fr(E)(n):=
\bigoplus_{\tT\in\tr_n}E(\tT)
\end{equation}
with the operad structure
on $\fr(E)$ given by the operation of `grafting'
trees. In the {\em symmetric case} we shall work with the
set of (abstract) trees with input edges indexed by finite ordered
sets. The formulas for $E(\tT)$ and $\fr(E)$ are similar but involve
also the symmetric group action, the details may be found
in~\cite{ginzburg-kapranov:preprint,getzler-jones:preprint}.
As mentioned earlier,
we try to discuss both the symmetric and nonsymmetric
cases simultaneously whenever possible.

In the special case when $E(m)=0$ for $m\not= 2$ the summation
in~(\ref{30}) reduces to the summation over the subset
$\tr^2_n\subset \tr_n$ consisting of {\em binary\/} trees, i.e.~trees
$\tT$ with $\val(v)= 2$ for any vertex $v\in \vert(\tT)$.
\eodst}\end{odstavec}

\begin{odstavec}{\rm {\bf --}
\label{37}
Let $\S = \prez ER$ be a quadratic operad as in~\ref{28}
and recall that $\S$ is
graded, $\S = \bigoplus_{i\geq 0}\S_i$.
Let $\trb_n$ denote the
set of 1-ternary binary $n$-trees, i.e. $n$-trees whose all vertices
have two incoming edges except exactly one which has three
incoming edges. Then we have for the collection
$\fr_{n-2,1}(E\op \tilde R)$ from~\ref{28}
\[
\fr_{n-2,1}(E\op \tilde R) = \bigoplus_{\tS \in \trb_{n+1}}
\tilde R_{\tS},
\]
where we denoted $\tilde R_{\tS}:= (E\op \tilde R)(\tS)$.

We may interpret the elements of $\tilde R_\tS$ as `multilinear'
colorings of $\tS$ such that
binary vertices are colored by
elements of $E$ and the only ternary
vertex is colored by an element of $R$.
Denote finally $R_\tS:= h(\tilde R_\tS)$. Then
$\S_n = \fr_n(E)/\sp(R_\tS;\
\tS\in \trb_{n+1})$. For the symmetric case
this type of description was given
in~\cite{ginzburg-kapranov:preprint}.
\eodst}\end{odstavec}

\begin{odstavec}{\rm {\bf --}
\label{45}
Let $U$ and $V$ be two collections and $E:= U\op V$. There
is an alternative way to describe the free operad $\fr(U,V):=
\fr(U\op V)$ resembling the description of the free associative
algebra $F(X\op Y)$ as the free product of $F(X)$ and $F(Y)$.
Let $\trwb$ be the set of 2-colored trees. This means that
the elements
of $\trwb$ are trees (planar in the nonsymmetric case, abstract in
the symmetric case) whose vertices are colored by two colors (`w' from
white, `b' from black). For $\tT\in\trwb$ let $\vert_w(\tT)$
(resp.~$\vert_b(\tT)$) denote the set of white (resp.~black) vertices
of $\tT$. Let $(U,V)(\tT)$ be the subset of $E(\tT)$ defined as
\[
(U,V)(\tT) :=
\bigotimes_{v\in\vert_w(\tT)}U(\val(v))\otimes
\bigotimes_{v\in\vert_b(\tT)}V(\val(v))
\]
Then we may define $\fr(U,V)(n)$ as
$\fr(U,V)(n):= \bigoplus_{\tT\in \trwb_n}(U,V)(\tT)$. If the
collections $U$ and $V$ are quadratic, the summation reduces to the
summation over the subset $\tr^{wb,2}$ of 2-colored {\em binary\/}
trees.
\eodst}\end{odstavec}

\begin{odstavec}{\rm {\bf --}
\label{35}
Recall that a tree is, by definition, an {\em oriented\/} graph.
Each edge $e$ has an output vertex $\out(e)$ and an input vertex
$\inp(e)$. This induces, by $\inp(e) \prec \out(e)$,
a partial order $\prec$ on the set $\vert(\tT)$ of vertices of $\tT$.
For a tree $\tT\in \tr^{wb}$ define $I(\tT)$ to be the number of
all couples $(v_1,v_2)$, $v_1 \in \vert_b(\tT)$ and $v_2 \in
\vert_w(\tT)$,
such that $v_2\prec v_1$.
\eodst}\end{odstavec}

By a differential graded (dg) collection we mean a collection $E =
\{E(n); n\geq 2\}$ such that each $E(n)$ is endowed with a
differential $d_E = d_E(n)$ which is, in the symmetric case, supposed
to commute with the symmetric group action.
For such a dg collection we define its (co)homology collection as
$H(E):= \{H(E(n),d_E(n)); n\geq 2\}$.
Let $U = \{(U(n), d_U(n));
n\geq 2\}$ and $V = \{(V(n), d_V(n)); n\geq 2\}$ be two dg
collections.
Define on $U\cd V$ the differential $d_{U\cd V}$ by
\[
d_{U\cd V}(\gamma(u;v_1,\ldots,v_k)) := \gamma(d_U(u);v_1,\ldots,v_k)
+ (-1)^{\deg(u)}\sum_{i=1}^k(-1)^{i+1}
\gamma(u;v_1,\ldots,d_V(v_i),\ldots,v_k)
\]
It can be easily verified that this formula introduces a monoidal
structure
on the category of dg collections. We formulate the following variant
of the K\"unneth theorem; recall that we assume the ground field $\bk$
to
be of characteristic zero.

\begin{proposition}
\label{38}
There exists a natural isomorphism of collections,
$H(U\cd V)\cong H(U)\cd H(V)$.
\end{proposition}

\noindent
{\bf Proof.}
In the nonsymmetric case we have the decomposition
\begin{equation}
\label{39}
(U\cd V)(m)=
\bigoplus(U\cd V)(l;k_1,\ldots,k_l),
\end{equation}
where $(U\cd V)(l;k_1,\ldots,k_l) :=
U(l)\ot V(k_1)\ot \cdots \ot V(k_l)$ and the summation is taken over
all $l\geq 2$ and $k_1+\cdots k_l=m$. The differential $d_{U\cd V}$
obviously preserves the decomposition and agrees on
$(U\cd V)(l;k_1,\ldots,k_l)$ with the usual tensor product
differential on $U(l)\ot V(k_1)\ot \cdots \ot V(k_l)$. The classical
K\"unneth theorem then gives the result.

For the symmetric case we have the same decomposition as
in~(\ref{39}),
but the summation is now taken over all $l\geq 2$ and
$k_1+\cdots +k_l=m$ {\em with\/} $k_1\leq k_2\leq \cdots\leq k_l$,
and $(U\cd V)(l;k_1,\ldots,k_l)$ is defined as
$\mbox{Ind}^{\Sigma_m}_{\Sigma_{k_1}
\times \cdots \times \Sigma_{k_l}}
(U(l)\ot V(k_1)\ot \cdots \ot V(k_l))$ where $\Sigma_{k_1}
\times \cdots \times \Sigma_{k_l}$ acts on $\Sigma_m$ via the
canonical inclusion and $\mbox{Ind}^{\Sigma_m}_{\Sigma_{k_1}
\times \cdots \times \Sigma_{k_l}}(-)$ denotes the induced action.
Since
$\mbox{char}(\bk)=0$, `the (co)homology commutes with finite group
actions' and we may use the same arguments as in the nonsymmetric
case.
\qed

\section{Distributive laws}

In this section we introduce the notion of a {\em distributive law\/}.
Let us begin with our `toy model' of associative algebras.
Suppose we have two quadratic associative algebras, $A =
\langle U,S\rangle$ and $B = \langle V,T\rangle$, and a
map $d: V\ot U \to U\ot V$. Let us denote $D:= \{v\ot u-d(v\ot u);\
v\ot u\in V\ot U\} \subset F_2(U\oplus V)$ and let $C:=
\langle U,V;S,D,T\rangle$ (= an abbreviation for $\langle U\oplus V;
S\oplus D\oplus T\rangle$). Observe that $C$ is bigraded, $C =
\bigoplus_{i,j\geq 0}C_{i,j}$, with $C_{1,0}=U$ and $C_{0,1}= V$,
the bigrading being induced by the natural
bigrading on $F(U,V):= F(U\oplus V)$.
We have a $\bk$-module map $\xi : A\ot B
\to C$ induced by the inclusion $F(U)\ot F(V)\subset F(U,V)$ which
preserves the bigrading, $\xi(A_i\ot B_j)\subset C_{i,j}$; let
$\xi_{i,j}:= \xi|_{A_i\ot B_j}$.

\begin{definition}
\label{15}
We say that $d$ defines a distributive law if
\begin{equation}
\label{0}
\xi_{i,j}: A_i\ot B_j\to C_{i,j} \mbox{ is an isomorphism for
$(i,j)\in\{(1,2),(2,1)\}$.}
\end{equation}
\end{definition}
To understand better the meaning of the condition~(\ref{0}) we
formulate the following lemma.

\begin{lemma}
\label{7}
The condition~(\ref{0}) is equivalent to
\begin{equation}
\label{4}
(\id \ot d)(d\ot \id)(V\ot S)\subset S\ot V \mbox{ and }
(d\ot \id)(\id \ot d)(T\ot U)\subset U\ot T,
\end{equation}
where $\id$ denotes the identity map.
\end{lemma}

\noindent
{\bf Proof.}
Let us consider the diagram
\[
[T^2(U)\ot V] \stackrel{(\id\ot d)}{\longleftarrow}
[U\ot V\ot U] \stackrel{(d\ot \id)}{\longleftarrow}
[V\ot T^2(U)].
\]
Then $C_{2,1}$ is obtained from the direct sum
\[
[T^2(U)\ot V] \oplus [U\ot V\ot U] \oplus [V\ot T^2(U)]
\]
by moding out $S\ot V\subset T^2(U)\ot V$,
$V\ot S\subset V\ot T^2(U)$, by identifying $x\in U\ot V\ot U$ with
$(\id\ot d)(x)\in T^2(U)\ot V$, and by identifying $y\in V\ot T^2(U)$
with $(d\ot \id)(y)\in U\ot V\ot U$. From this we see immediately
that $\xi_{2,1}$ is an isomorphism if and only if
$(\id \ot d)(d\ot \id)(V\ot S)\subset S\ot V$, which is the first
relation of~(\ref{4}).
Similarly we may show that $\xi_{1,2}$ is an
isomorphism if and only if the second relation of~(\ref{4}) is
satisfied.
\qed

The following theorem shows that the map $\xi_{i,j}$ is an
isomorphism for $(i,j)\in \{(2,1),(1,2)\}$ if and only if it is an
isomorphism for an arbitrary couple $(i,j)$.

\begin{theorem}
\label{8}
Suppose $d$ is a distributive law. Then the map
\[
\xi_{i,j}: A_i\ot B_j\to C_{i,j}
\]
is an isomorphism for all $(i,j)$.
\end{theorem}

\noindent
{\bf Proof.}
Let us fix $i,j\geq 0$. The fact that $\xi_{i,j}$ is an epimorphism
is clear. The map $\xi_{i,j}$ is a monomorphism if, for any
$a\in T^i(U)\ot T^j(V)$, $a=0 \mod (S,D,T)$ in $F_{i,j}(U,V)$ implies
$a=0 \mod (S,T)$ in $T^i(U)\ot T^j(V)\subset F_{i,j}(U,V)$.

Let us introduce the following terminology. We say that an element
$b \in F_{i,j}(U,V)$ is a monomial if it is of the form
$b =b_1\ot\cdots \ot b_{i+j}$ with $b_k \in U$
or $b_k\in V$ for $1\leq k\leq i+j$. For a monomial $b$ define $I(b)$
to be the number of inverses in $b$, i.e. the number of couples
$(b_k,b_l)$ such that $b_k\in V$, $b_l\in U$ and $k < l$. Observe
that monomials in $F_{i,j}(U,V)$ linearly
generate $F_{i,j}(U,V)$.

Let $b = b_1\ot \cdots \ot b_{i+j}\in F_{i,j}(U,V)$ be a monomial.
We say that a number $s$, $1\leq s\leq i+j-1$,
is $b$-admissible if $b_s\in V$
and $b_{s+1}\in U$. We say that $s$ is $x$-admissible, for
$x\in F_{i,j}(U,V)$, if $x = \sum_{\omega\in\Omega}b_\omega$ for some
monomials $b_\omega\in F_{i,j}(U,V)$
such that $s$ is $b_\omega$-admissible for any
$\omega\in \Omega$.
Let us denote $d(s):= \id^{s-1}\ot d\ot \id^{i+j-s-1}$.
Then $a=0 \mod (S,D,T)$
means the existence of a finite set $K$, elements
$a_\kappa\in F_{i,j}(U,V)$ and $a_\kappa$-admissible numbers
$s_\kappa$,
$\kappa \in K$, such that
\begin{equation}
\label{1}
a =
\sum_{\kappa \in K}
(a_\kappa -d(s_\kappa)(a_\kappa))
\mod (S,T).
\end{equation}
We say that $a=0 \modn N $, for
$N\geq 0$, if $\mbox{\rm max}\{I(a_\kappa);\ \kappa\in K\}\leq N$.
Obviously, $a=0 \modn0$ if and only if $a=0 \mod (S,T)$ and $a=0
\mod (S,D,T)$ if and only if $a=0 \modn N$ for some $N$. This means
that it is enough to prove the following lemma.

\begin{lemma}
\label{3}
If $a=0 \modn N$ for some $N\geq 1$, then $a=0 \modn {N-1}$.
\end{lemma}

Let us prove first the following equation. Let $b$ be a monomial,
$I(b)=N$, and let $s_1,s_2$ be two $b$-admissible numbers. Then
\begin{equation}
\label{2}
b-d(s_1)(b) = b-d(s_2)(b) \modn {N-1}.
\end{equation}
The equation follows immediately from the obvious commutativity
$d(s_1)d(s_2)= d(s_2)d(s_1)$ which implies that
\begin{equation}
\label{14}
b-d(s_1)(b) = b-d(s_2)(b) + (d(s_2)(b)-d(s_1)d(s_2)(b))-
(d(s_1)(b)-d(s_2)d(s_1)(b))
\end{equation}
which, together with the observation that $d(s_k)(b)$ is, for
$k=1,2$, a sum of monomials with $I < N$, finishes the proof of the
equation.
Relation~(\ref{2}) says, loosely speaking, that the
sequence $\{s_\kappa\}_{\kappa \in K}$ can be replaced in the proof of
Lemma~\ref{3} by any other sequence of admissible numbers.

Let us denote $K_N:= \{\kappa;\ I(a_\kappa)=N\}$. Obviously $a_N :=
\sum_{\kappa \in K_N}a_\kappa = 0 \mod (S,T)$. This means that $a_N =
\sum_{1\leq t\leq i+j-1}(a^S_t+a^T_t)$, for some $a_t^S
\in (U\op V)^{t-1}\ot S \ot (U\op V)^{i+j-t-1}$ and $a_t^T
\in (U\op V)^{t-1}\ot T \ot (U\op V)^{i+j-t-1}$. For a fixed $t$
consider the element $a^S_t$. Suppose that there exists an
$a^S_t$-admissible number $s$, $s\not= t-1$. Then obviously
$d(s)(a^S_t) \in (S)$ and we may subtract $a^S_t - d(s)(a^S_t)$ from
the right-hand side of~(\ref{1}). Suppose this is not the case, i.e.
that the only $a^S_t$-admissible number is $s=t-1$.
Then clearly $s+1$
is $d(s)(a^S_t)$-admissible and
\[
a^S_t -d(s)(a^S_t) + d(s)(a^S_t) - d(s+1)d(s)(a^S_t) \in (S)
\]
by the first equation of~(\ref{4}).
We may again subtract the above expression from the right-hand side
of~(\ref{1}). In both cases we got rid of $a^S_t$ and we may repeat
the process for all $t$, $1\leq t\leq i+j-1$. The terms $a^T_t$
can be removed in exactly the same way, so we end up with $a_N=0$.
This finishes the proof of Lemma~\ref{3}.
\qed

\vskip-7mm
\hphantom{p}
\begin{odstavec}{\rm{\bf --}
\label{40}
Let us introduce distributive laws for operads.
Let $\A = \langle U;S\rangle$ and $\B = \langle V;T\rangle$ be two
quadratic operads. Let $V\b U$ denote the subcollection of $\fr(U,V)$
generated by elements of the form $\gamma(v;u,1)$ or $\gamma(v;1,u)$,
$u\in U$ and $v\in V$. Clearly $(V\b U)(m)=0$ for $m\not= 3$. The
notation $U\b V$ has the obvious similar meaning.
Suppose we have a map $d: V\b U \to U\b V$
of collections and let $D := \{z-d(z);\ z\in V\b U\}
\subset \fr(U,V)(3)$ and $\C:= \langle U,V;S,D,T\rangle$ (= an
abbreviation for $\langle U\oplus V; S\oplus D \oplus T\rangle$).
As in the case of associative algebras, the inclusion
$\fr(U)\cd \fr(V)\subset \fr(U,V)$, where $\cd$ is the operation from
Proposition~\ref{27}, induces a map $\xi :\A\cd \B \to \C$ of
collections. The collection $\fr(U,V)$ is bigraded (see~\ref{26}) and
the
relations $S,T$ and $D$ obviously preserve
this bigrading, hence the
operad $\C$ is naturally bigraded as well.
Also the collection $\A \cd \B$ is bigraded:
$(\A\cd\B)_{i,j}$ is generated by elements of the form
$\gamma(a;b_1,\ldots,b_{i+1})$, $a\in \A_i$ (= $\A(i+1)$) and
$b_k\in \B_{j_k}$ (= $\B(j_k+1)$) for $1\leq k \leq i+1$ and
$j_1+\cdots+j_{i+1}=j$. We write
more suggestively $\A_i\cd \B_j$ instead of $(\A\cd\B)_{i,j}$,
abusing the notation a bit.
\eodst}\end{odstavec}

As in the case of associative algebras, the map $\xi$ preserves the
bigrading, $\xi(\A_i\cd \B_j)\subset \C_{i,j}$, let $\xi_{i,j}:=
\xi|_{\A_i\cd \B_j}$. We have the
following analog of Definition~\ref{15}.

\begin{definition}
\label{16}
The map $d: V\b U \to U\b V$ defines a distributive law
if the following condition is satisfied:
\begin{equation}
\label{6}
\xi_{i,j}: \A_i\cd \B_j\to \C_{i,j} \mbox{ is an isomorphism for
$(i,j)\in\{(1,2),(2,1)\}$.}
\end{equation}
\end{definition}

There exists a way of rewriting this condition into a more
explicit form as it was
done in Lemma~\ref{7} for associative algebras, but the resulting
formulas are much more complicated and we postpone the discussion of
this to Section~5.
The main result of this section is the following analog of
Theorem~\ref{8}.

\begin{theorem}
\label{13}
Suppose $d$ is a distributive law. Then the map
$\xi_{i,j}: \A_i\cd \B_j\to \C_{i,j}$
is an isomorphism for all $(i,j)$.
\end{theorem}

It is clear that $\xi_{i,j}$ is an epimorphism.
The proof that $\xi_{i,j}$ is a monomorphism will occupy the
rest of this section. We must prove that
$a\in \fr_i(U)\cd \fr_j(V)\subset \fr_{i,j}(U,V)$
is zero mod $(S,D,T)$ if and only if it
is zero mod $(S,T)$.

Let $\tr^{wb,2}(i,j)$ denote the subset of $\tr^{wb,2}$ consisting
of threes having exactly $i$ white and $j$ black vertices; we observe
that $\fr_{i,j}(U,V)= \bigoplus_{\tT\in \tr^{wb,2}(i,j)}(U,V)(\tT)$,
see~\ref{45} for the notation.

Let $\tT\in \tr^{wb,2}(i,j)$, $v\in \vert(\tT)$ and $\epsilon$
be an input edge of $v$. We say that the couple $(v,\epsilon)$ is
$\tT$-{\em admissible\/} if $v\in \vert_b(\tT)$ and
$\inp(\epsilon)\in \vert_w(\tT)$. Sometimes we also say that
$(v,\epsilon)$ is $b$-admissible if $b\in (U,V)(\tT)$ and if
$(v,\epsilon)$ is $\tT$-admissible.

Let us suppose that
$(v,\epsilon)$ is $\tT$-admissible. Let us denote by $\tS$ the
minimal binary subtree of $\tT$ containing $v$ and $w:=
\inp(\epsilon)$. Clearly $\tS \in \tr^{wb,2}_3(1,1)$ and $I(\tS)= 1$
(for the definition of $I(\tS)$ see~\ref{35}).
Let $b\in (U,V)(\tT)$ be of the form
$b=\bigotimes_{v\in \vert_w(\tT)}u_v \ot
\bigotimes_{v\in \vert_b(\tT)}v_v$ for some $u_v\in U$ and $v_v\in V$.
We call elements of this form monomials and we observe that monomials
generate $(U,V)(\tT)$. For a monomial $b$ as above let
$\overline b_{\tS}\in (U,V)(\tS) \subset V\b U$ be defined as
$\overline b_{\tS} := \bigotimes_{v\in \vert_w(\tS)}u_v \ot
\bigotimes_{v\in \vert_b(\tS)}v_v$ (observe however that both
$\vert_w(\tS)$ and $\vert_b(\tS)$ consist of one element). Let
$\Xi := \{\tR\in \tr_3^{wb,2}(1,1); I(\tR)=0\}$, we note that $\Xi$
consist of exactly two (resp.~three) trees in the symmetric
(resp.~nonsymmetric) case. Then $d(\overline b_{\tS})=
\sum_{\tR\in \Xi}\overline b'_{\tR}$ for some
$\overline b'_{\tR}\in (U,V)(\tR)\subset U\bullet V$.
Let $\tT_{\tR}$ denote the tree obtained from $\tT$ by replacing the
subtree $\tS$ by $\tR$; observe that $I(\tT_{\tR})< I(\tT)$.
Let finally $b'_{\tR}\in (U,V)(\tT_{\tR})$ be, for $\tR\in \Xi$, an
element obtained by substituting $\overline b'_{\tR}$ to $b$ at the
vertices of $\tR$. Let us define $d(v,\epsilon)(b):=
\sum_{\tR\in \Xi}b'_{\tR} \in \fr_{i,j}(U,V)$ and let us extend this
definition linearly (and equivariantly in the symmetric case)
to the whole $(U,V)(\tT)$. Loosely speaking,
$d(v,\epsilon)(b)$ is obtained from $b$ by making the `surgery'
prescribed by the distributive law at the couple $(v,\epsilon)$.

Under the above notation the condition
$a = 0 \mod (S,D,T)$ means that there
exist a finite set $K$, trees $\tT_\kappa \in \tr^{wb,2}(i,j)$,
elements $a_\kappa \in (U,V)(\tT_\kappa)$, vertices
$v_\kappa \in \vert(\tT_\kappa)$ and
edges $\epsilon_\kappa$ such that $(v_\kappa,\epsilon_\kappa)$ is
$\tT_k$-admissible and
\begin{equation}
\label{9}
a= \sum_{\kappa\in K}(a_\kappa -
d(v_\kappa,\epsilon_\kappa)(a_\kappa)) \mod (S,T).
\end{equation}
Similarly as in the associative algebra case we say that $a = 0 \modn
N$ if
$\mbox{\rm max}\{I(\tT_\kappa);\ \kappa \in K\}\leq N$. It is again
enough to prove the following analog of Lemma~\ref{3}.

\begin{lemma}
\label{10}
If $a=0 \modn N$ for some $N\geq 1$, then $a=0 \modn {N-1}$.
\end{lemma}

We would need also an analog of relation~(\ref{2}). Since the proof
will be much more complicated than in the associative algebra case,
we formulate it as a separate statement.

\begin{lemma}
\label{11}
Let $b\in (U,V)(\tT)$, $I(\tT)= N$ and suppose $(v_1,\epsilon_1)$ and
$(v_2,\epsilon_2)$ are two $b$-admissible couples. Then
\begin{equation}
\label{12}
b-d(v_1,\epsilon_1)(b) = b-d(v_2,\epsilon_2)(b) \modn{N-1}.
\end{equation}
\end{lemma}

\noindent
{\bf Proof.}
The lemma is trivially satisfied for
$(v_1,\epsilon_1)=(v_2,\epsilon_2)$, so suppose
$(v_1,\epsilon_1)\not=(v_2,\epsilon_2)$.

Let us discuss the case $v_1\not= v_2$ first. Then
$(v_2,\epsilon_2)$ is admissible for each monomial in
$d(v_1,\epsilon_1)(b)$ because the rearrangements made by
$d(v_1,\epsilon_1)$ does not change the vertex $v_2$ and the edge
$\epsilon_2$. Similarly, $(v_1,\epsilon_1)$ is
admissible for each monomial in
$d(v_2,\epsilon_2)(b)$. We also notice that
$d(v_1,\epsilon_1)d(v_2,\epsilon_2)(b)=
d(v_2,\epsilon_2)d(v_1,\epsilon_1)(b)$.
We therefore have the following analog
of~(\ref{14})
\[
(\id - d(v_1,\epsilon_1))(b)=(\id - d(v_2,\epsilon_2))(b)
+(\id - d(v_1,\epsilon_1))d(v_2,\epsilon_2)(b) -
(\id - d(v_2,\epsilon_2))d(v_1,\epsilon_1)(b),
\]
which implies~(\ref{12}).

Suppose $v_1=v_2 =:v$, then, of course, $\epsilon_1\not=\epsilon_2$.
Let $w_1=\inp(\epsilon_1)$ and $w_2=\inp(\epsilon_2)$. Let $\tT'$ be
the smallest binary subtree of $\tT$ containing
$v,w_1$ and $w_2$. Obviously $\tT' \in \tr^{wb,2}_4(2,1)$ and
$I(\tT')= 2$.
Because the surgery made by both
$d(v_1,\epsilon_1)$ and $d(v_2,\epsilon_2)$
takes place inside $\tT'$, we may suppose that in fact $\tT' =
\tT \in \tr^{wb,2}_4(2,1)$. Then $b\in \fr_1(V)\cd \fr_2(U)$ and
$d(v_1,\epsilon_1)(b) = \sum_{\tR\in \Omega_0}\overline b_{\tR}+
\sum_{\tR\in \Omega_1}\overline b_{\tR}$, where
$\Omega_i= \{\tR\in \tr^{wb,2}_4(2,1); I(\tR)= i\}$,
$\overline b_{\tR}\in (U,V)(\tR)$, $i=1,2$. Applying on the summands
of the second sum the distributive law once again (which can be done
in exactly one way as there is exactly one admissible couple for any
$\tR \in \Omega_1$) we obtain
some $b' \in \fr_2(U)\cd \fr_1(V)$, $b' = b \mod (S,D,T)$.
By the same process with the
r\^oles of $d(v_1,\epsilon_1)$ and $d(v_2,\epsilon_2)$ interchanged
we construct
another element $b'' \in \fr_2(U)\cd \fr_1(V)$. But $b' = b''
\mod (S,T)$, by~(\ref{6}) with $(i,j)= (2,1)$. This finishes the
proof of Lemma~\ref{11}.
\qed

Relation~(\ref{12}) says that the concrete values of the
couples $(v_\kappa,\epsilon_\kappa)$ in~(\ref{9}) are not substantial.
Let us come back to
the proof of Lemma~\ref{10}. We introduce the following clumsy
notation. By $\tr^{wb,2,w3_1}$ we denote the set of 2-colored binary
1-ternary trees such that the ternary vertex is white. The notation
$\tr^{wb,2,b3_1}$ will have the obvious similar meaning.

Let $K_N := \{\kappa \in K;\
I(\tT_{\kappa}) =N\}$. Then necessarily
$a_N := \sum_{\kappa \in K_N}a_\kappa = 0
\mod (S,T)$ which means that $a_N =
\sum_{\omega\in \Omega} a_\omega^S+\sum_{\delta\in \Delta}
a_\delta^T$, where $a_\omega^S$ is an
element of $S_{\tS_\omega}$, $\tS_\omega \in \tr^{wb,2,w3_1}$ and,
similarly, $a_\delta^T$ is an
element of $T_{\tS_\delta}$, $\tS_\delta \in \tr^{wb,2,b3_1}$;
see~\ref{37} for the notation.

Let us discuss the term $a^S_\omega$ for a
fixed $\omega\in \Omega$ first. Suppose there
exists a black vertex $v\in \vert_b(\tS_\omega)$ and an edge
$\epsilon$ with $\out(\epsilon)= v$ such that $w:=\inp(\epsilon)$
is white and binary.
Then obviously $d(v,\epsilon)(a^S_\omega)$ makes sense and
$d(v,\epsilon)(a^S_\omega) = 0 \mod (S)$ since $d(v,\epsilon)$ does
not change the ternary vertex of $\tS_\omega$. We can delete
$a^S_\omega - d(v,\epsilon)(a^S_\omega)$ from the right-hand side
of~(\ref{9}).

Suppose that the only edge $\epsilon$ of $\tS_\omega$ such
that $\out(\epsilon)$ is black is the output edge of the ternary
white vertex $v_3$. Let us pick this edge $\epsilon$ and denote $v :=
\out(\epsilon)$ its black output vertex. Let
$\tR \in \tr^{wb,2,w3_1}_4(1,1)$
be the minimal tree containing $v_3$ and $v$. Using the same locality
argument as before we may suppose that in fact $\tS_\omega = \tR$.
Then $a^S_\omega \in \fr_{2,1}(U,V)$ and
we may replace $a^S_\omega$ modulo $(D)$ by some
${a'}^S_\omega\in \fr_2(U)\cd \fr_1(V)$. We infer
form~(\ref{6}) with $(i,j)= (2,1)$
that ${a'}^S_\omega = 0 \mod (S)$, so we may delete $a^S_\omega$
from the right-hand side of~(\ref{9}). The discussion of the second
type terms is similar.
\qed

\section{Distributive laws and triples}

In this paragraph we discuss the relation of our
definitions with the triple definition
of a distributive law as it was originally given by J.~Beck
in~\cite{beck:LNM80}.
We show that both definitions coincide; the hard part of this
statement has been in fact already proven in the previous section
(Theorem~\ref{13}).

Recall~\cite{getzler-jones:preprint}
that each operad $\S$ generates a triple
$\sfT=(\sfT,\mut,\etat)$ on the category of vector spaces as follows.
The functor $\sfT$ is defined by $\sfT(V) :=
\bigoplus_{n\geq 1}(\S(n)\ot T^n(V))$
in the nonsymmetric case and
$\sfT(V) := \bigoplus_{n\geq 1}(\S(n)\ot T^n(V))_{\Sigma_n}$,
where $(-)_{\Sigma_n}$ denotes the coinvariants of the symmetric
group action on the product $\S(n)\ot T^n(V))$ given by the operad
action on the first factor and by permuting the variables of the
second factor, in the symmetric
case. The transformation $\etat :\id \to \sfT$ is defined by the
inclusion $V = \S(1)\ot V \subset \sfT(V)$ (we suppose here as always
that $\S(1)= \bk$; this condition is automatically
satisfied for quadratic operads). To define the transformation $\mut$
observe first that $\sfT \sfT (V)=
\bigoplus_{n\geq 1}((\S \cd\S)(n)\ot T^n(V))$ (in the symmetric case
we must take the coinvariants), the transformation
$\mut :\sfT\sfT \to \sfT$ is then induced by the operad
multiplication $\gamma_\S : \S \cd \S \to \S$. This construction has
the property that algebras over the operad $\S$ are the same as
algebras over the triple $\sfT$.

Let $\sfS= (\sfS,\mus,\etas)$ and $\sfT = (\sfT,\mut,\etat)$
be two triples. Let us recall the following definition
of~\cite[page~120]{beck:LNM80}.

\begin{definition}
\label{31}
A distributive law of $\sfS$ over $\sfT$ is a natural
transformation $\ell: \sfT\sfS \to \sfS\sfT$ having the following
properties.
\begin{eqnarray}
\label{32}
&\ell \c \sfT\etas = \etas \sfT \mbox{ and }
\ell \c \etat\sfS = \sfS\etat&
\\
\label{33}
&\mus \sfT \c \sfS \ell \c \ell \sfS = \ell \c \sfT\mus&
\\
\label{34}
&\sfS \mut \c \ell \sfT \c \sfT\ell = \ell \c \mut \sfS&
\end{eqnarray}
\end{definition}

Let $\C = \langle U,V; S,D,T\rangle$ be an operad with a distributive
law in the sense of our Definition~\ref{16}, $\A := \langle
U,S\rangle$ and
$\B := \langle V,T\rangle$. Let $\sfS = (\sfS,\mus,\etas)$
(resp~$\sfT = (\sfT,\mut,\etat))$ be the triple
associated to the operad $\A$ (resp.~$\B$). We
have, in the nonsymmetric case,
\[
\sfS\sfT(V) := \bigoplus_{n\geq 1}((\A\cd \B)(n)\ot T^n(V))
\mbox{ and }
\sfT\sfS(V) := \bigoplus_{n\geq 1}((\B\cd \C)(n)\ot T^n(V)),
\]
while obvious similar formulas hold, after
taking the coinvariants, also in the symmetric case.
There is a natural map of collections $\Delta :\B\cd \A
\to \A\cd \B$ given as the composition
\[
\B\cd \A \stackrel{\chi}{\longrightarrow}
\C \stackrel{\xi^{-1}}{\longrightarrow} \A\cd \B
\]
where $\chi$ is induced by the inclusion
$\fr(V)\cd \fr(U)\subset \fr(U,V)$ and $\xi :\A\cd \B \to \C$ is the
map introduced in~\ref{40}. We used the nontrivial fact
that the map $\xi$ is an isomorphism (Theorem~\ref{13}).
The map $\Delta :\B\cd \A
\to \A\cd \B$ then induces, by $\ell (x\ot y):= \Delta (x)\ot y$,
$x\ot y \in (\B\cd\A)(n)\ot T^n(V)$, a natural transformation $\ell:
\sfT\sfS \to \sfS\sfT$.

\begin{theorem}
\label{48}
Under the notation above, the transformation $\ell:
\sfT\sfS \to \sfS\sfT$ is a distributive law in the sense
of Definition~\ref{31}.
\end{theorem}

\noindent
{\bf Proof.}
For two collections $U$ and $V$ let $\iota_{U,U\cd V}:U\to U\cd V$
and $\iota_{V,U\cd V}: V\to U\cd V$ denote the inclusions.
Condition~(\ref{32}) is then equivalent to
\[
\Delta \circ \iota_{\B,\B\cd\A}= \iota_{\B,\A\cd B}
\mbox{ and }
\Delta \circ \iota_{\A,\B\cd\A}= \iota_{\A,\A\cd B}
\]
which is immediately seen. Condition~(\ref{33}) is equivalent to
\[
(\mus\cd\id)\circ(\id\cd\Delta)\circ(\Delta \cd \id)=
\Delta \circ(\id\cd\gamma_{\A}),
\]
while~(\ref{34}) is equivalent to
\[
(\id\cd\mut)\circ(\Delta \cd \id)\circ(\id\cd\Delta)=
\Delta \circ(\gamma_{\B}\cd \id).
\]
We may safely leave the verification of these two equations to the
reader.
\qed

\section{Explicit computation and examples}

In this section we aim to discuss the condition~(\ref{6}) of
Definition~\ref{16} and give also some explicit examples of
operads with a distributive law and algebras over them. We keep the
notation introduced in the previous sections.
Let us discuss the
nonsymmetric case first.

By definition, a distributive law is given by a map $d: V\b U \to
U\b V$. Because $V\b U =V\c_1 U\op V\c_2 U$ and $U\b V =U\c_1
V\op U\c_2 V$, such a map is given by a $2\times 2$ matrix
$(\alpha_{i,j})$ with $\alpha_{i,j}: V\c_i U \to U\c_j V$ a linear
map, $i,j= 1,2$.

\begin{proposition}
\label{17}
The matrix $(\alpha_{i,j})$ defines a distributive law if and only if
the following conditions are satisfied.
\begin{enumerate}
\item
$\aa11 = \aa22 = 0$
\item
$(\id_U\c_2\aa21)(\aa12 \c_3 \id_U) =
(\id_U\c_1\aa12)(\aa21 \c_1 \id_U) \mod (S)$
\item
Let $v\in V$, $s_1 \in U\c_1U$ and $s_2 \in U\c_2U$ be such that
$s_1\op s_2\in S \subset \fr(U)(3) = U\c_1U\op U\c_2U$. Then
\begin{eqnarray*}
(\aa12 \c_1\id_U)(v\c_1 s_1)
&=& (\id_U\c_2\aa12)(\aa12\c_2\id_U)(v\c_1 s_2) \mod (S), \mbox{ and}
\\
(\aa21 \c_3\id_U)(v\c_2 s_2)
&=& (\id_U\c_1\aa21)(\aa21\c_1\id_U)(v\c_2 s_1)\mod (S).
\end{eqnarray*}
\item
Let $u\in U$, $t_1 \in V\c_1V$ and $t_2 \in V\c_2V$ be such that
$t_1\op t_2\in T$. Then
\begin{eqnarray*}
(\aa12 \c_2 \id_V)(\id_V\c_1 \aa12)(t_1\c_1 u)
&=& (\aa12 \c_3 \id_V)(t_2\c_1 u) \mod (T),
\\
(\aa12\c_1 \id_V)(\id_V \c_1 \aa21)(t_1\c_2 u)
&=&
(\aa21\c_3 \id_V)(\id_V \c_2 \aa12)(t_2\c_2 u) \mod (T), \mbox{ and}
\\
(\aa21 \c_1 \id_V)(t_1\c_3 u)
&=&
(\aa21\c_2\id_V)(\id_V\c_2 \aa21)(t_2\c_3 u) \mod (T).
\end{eqnarray*}
\end{enumerate}
\end{proposition}

We leave the verification, which is technical but absolutely
straightforward, to the reader.
Let us make some comments on the meaning of the equations above.
Take, for example, the second one. Both sides act on $\gamma(V;U,U)$,
let $x \in \gamma(V;U,U)$. Because $\gamma(V;U,U) = (V\c_1 U)\c_3 U$,
the expression $(\aa21 \c_3 \id_U)(x)$ makes sense and it is an
element of $(U\c_2 V)\c_3U$. But $\mbox{$(U\c_2 V)\c_3U$}
= U\c_2 (V\c_2 U)$
and $(\id_U\c_2\aa21)(\aa12 \c_3 \id_U)(x)$ again makes sense and it
is an element of $U\c_2 (U\c_1 V)\subset \fr_{2,1}(U,V)$. Similarly,
the right-hand side applied on $x$ is another element of
$\fr_{2,1}(U,V)$ and this equation says that these elements are the
same in $\fr(U,V)/(S)$. The remaining conditions have a similar
meaning.

\begin{example}{\rm\
\label{49}
Let us discuss a special type of solutions of the conditions in
Proposition~\ref{17}. Notice first that we have the canonical
identifications of vector spaces
$U\c_i V = U\ot V$ and $V\c_i U = V\ot U$, $i=1,2$. Using this
identification, define $\aa12 (v\ot u)= \aa21 (v\ot u):=u\ot v$.
Similarly, we have the identifications $U\c_1 U = U\ot U$ by
$\gamma(u_1;u_2,1)= u_1\ot u_2$ and $U\c_2 U = U\ot U$ by
$\gamma(u_1;1,u_2)= u_1\ot u_2$. In the same vain, we have the
identifications $V\c_i V = V\ot V$, $i=1,2$. Under these
identifications, $S\subset U\c_1 U \op U\c_2 U$ can be considered as
a subset of $U\ot U \op U\ot U$ and, similarly, $T$ can be
interpreted as a subset of $V\ot V\op V\ot V$.

The first equation of Proposition~\ref{17} is satisfied trivially.
The second condition means, for $(\alpha_{i,j})$ as above,
that $S\subset U\ot U\op U\ot U$ contains all `diagonal'
elements $u_1\ot u_2 \op (-u_2\ot u_1)$, $u_1,u_2\in U$.
The third condition is satisfied automatically while the last
one says that $T = \{z\op (-s(z));\ z\in Z\}$,
for a subspace $Z\subset V\ot V$, where $s:V\ot V\to V\ot V$ denotes
the `switch', $s(v_1\ot v_2)= v_2\ot v_1$.

An algebra over this type of operad is a
vector space $V$ with two sets of bilinear
operations, $\{(-,-)_i\}_{1\leq i\leq k}$ and
$\{\langle-,-\rangle_j\}_{1\leq j\leq l}$, such that
\begin{enumerate}
\item
The operations $(-,-)_i$ are mutually associative,
\[
((a,b)_i,c)_j = (a,(b,c)_j)_i,\ 1\leq i,j\leq k,
\]
and some more axioms may be imposed.
\item
The following distributivity laws are satisfied
\[
\langle(a,b)_i,c\rangle_j = (a,\langle b,c \rangle_j)_i
\ \mbox{ and }\
\langle a,(b,c)_i\rangle_j = (\langle a,b \rangle_j,c)_j,\ 1\leq
i\leq k,\ 1\leq j\leq l.
\]
\item
There are scalars $a^n_{ij}\in \bk$, $1\leq n\leq N$, such that
\[
\sum_{1\leq i,j\leq l} a^n_{ij}\{\langle\langle
a,b\rangle_i,c\rangle_j
-\langle a,\langle b,c \rangle_j\rangle_i\} = 0, \mbox{ for }
1\leq n \leq N.
\]
\end{enumerate}
}\end{example}

\begin{example}{\rm\
\label{47}
In this example we aim to discuss operads $\C = \langle U,V;
S,D,T\rangle$ with a distributive law such that $U\cong V\cong \bk$.
As $V\c_1 U\cong V\c_2 U \cong \bk$, the matrix $(\aa ij)$ of
Proposition~\ref{17} reduces (taking into the account the first
condition) to a couple $(\aa12,\aa21)$ of elements of $\bk$.

Let $\S = \prez ER$ be a nonsymmetric operad
with $E\cong \bk$. Let us fix a nonzero
element $e\in E$. Let $\whk:= \bk \op \bk \cup \{(\infty,\infty)\}$
(disjoint union of $\bk\op\bk$ with an `abstract' point $\ii$) and
define an action of $\bk_0:= \bk \setminus \{0\}$ on
$\whk$ by $\alpha(x,y):= (\alpha x,\alpha y)$ for
$(x,y)\in \bk\op\bk$ and $\alpha \ii :=
\ii$, $\alpha \in \bk_0$. For $(a,b)\in \whk$ let $[a,b]$ denote the
class of $(a,b)$ in $\whkmod$. Let $R_{[a,b]}\subset \fr(E)(3)$ be
the subspace defined by
\[
R_{[a,b]}:=\left\{
\begin{array}{ll}
\mbox{Span}(a(e\c_1e)- b(e\c_2e));
&\mbox{ for $(a,b)\in \bk\op \bk$, and}
\\
\fr(E);&\mbox{ for $(a,b)= \ii$.}
\end{array}
\right.%\}
\]
It is immediate to see that the correspondence
$[a,b]\leftrightarrow \prez E{R_{[a,b]}}$ is an one-to-one
correspondence between points of $\whkmod$ and isomorphism classes of
operads $\S = \prez ER$ with $E\cong \bk$. We also note that this
correspondence does not depend on the choice of $e\in E$.

Summing up the above remarks we see that our operad $\C = \langle U,V;
S,D,T\rangle$ is determined by a triple
$([a_U,b_U],(\aa12,\aa21),[a_V,b_V])$, where $(\aa12,\aa21)$ are the
matrix elements of the distributive law and $[a_U,b_U]$
(resp.~$[a_V,b_V]$) are elements of $\whkmod$ corresponding to $\A =
\prez US$ (resp.~$\B= \prez VT$). Making a detailed analysis of the
conditions of Proposition~\ref{17} we may see that the triple
$([a_U,b_U],(\aa12,\aa21),[a_V,b_V])$ determines an operad with a
distributive law if and only if (at least) one of the following
conditions is satisfied
\begin{enumerate}
\item[(i)]
$(\aa12,\aa21)=(0,0)$
\item[(ii)]
$(\aa12=0)$ and \hfill\break
$(\aa21=1) \mbox{ or }
((a_U\cdot b_U=0 \mbox{ or } (a_U,b_U)= \ii)
\mbox{ and }((a_V\cdot b_V=0 \mbox{ or } (a_V,b_V)= \ii))$
\item[(iii)]
$(\aa21=0)$ and \hfill\break
$(\aa12=1) \mbox{ or }
((a_U\cdot b_U=0 \mbox{ or } (a_U,b_U)= \ii)
\mbox{ and }((a_V\cdot b_V=0 \mbox{ or } (a_V,b_V)= \ii))$
\item[(iv)]
$([a_U,b_U]=[1,1]\mbox{ or }[a_U,b_U]= [\infty,\infty])
\mbox{ and }
(a_V=b_V) \mbox{ and }\hfill\break
(\aa12\!=\!1 \mbox{ and } \aa21\! =\!1)
\mbox{ or }((a_U,b_U)\!=\!\ii)
\mbox{ and }((a_V,b_V)\!=\!(0,0) \mbox{ or }(a_V,b_V)\!=\!\ii)$
\end{enumerate}
An important value of the triple
$([a_U,b_U],(\aa12,\aa21),[a_V,b_V])$ which
satisfies the last condition is
$([1,1],(1,1),[0,0])$. An algebra over the corresponding operad
is a vector space $V$
with an associative multiplication $\cdot$ and a bilinear operation
$\langle -,-\rangle$ such that
\[
\langle a\cdot b,c\rangle = a\cdot \langle b,c\rangle, \mbox{ and }
\langle a,b\cdot c\rangle = \langle a,b\rangle \cdot c.
\]
This algebra is a {\em nonsymmetric analog\/} of a Poisson algebra.
It is, moreover, of the type discussed in the previous
example. The computation above gives also very strange examples of
algebras with a distributive law. For example, an algebra over the
operad given by $([1,0],(3,0),(0,1))$ consists of a vector space $V$
and two bilinear operations $\cdot$ and $\langle -,-\rangle$ such
that, for each $a,b,c\in V$,
\[
(a\cdot b)\cdot c=0,\ \langle a\cdot b,c\rangle = 3a
\cdot\langle b,c\rangle,\ \langle a,b\cdot c\rangle = 0 \mbox{ and }
\langle a,\langle b,c \rangle\rangle =0.
\]
}\end{example}

Let us discuss the symmetric case. The space $V\b U$ now decomposes
as $V\b U = (V\b U)_1 \op (V\b U)_2\op (V\b U)_3$ with
\[
(V\b U)_1 := V\c_1U,\ (V\b U)_2:= V\c_2U \mbox{ and } (V\b U)_3 :=
(V\c_1U)(1\otimes \sigma),
\]
with $\sigma =$ the generator of $\Sigma_2$.
We have, of course, the similar decomposition also for $U\b V$, hence
$d:V\b U\to U\b V$ is given by a $3\times 3$-matrix $(\beta_{i,j})$,
$\beta_{i,j}: (V\b U)_i \to (U\b V)_j$. Since $d$ is, by definition,
a $\Sigma_3$-equivariant map and $(V\b U)_1$ generates $V\b U$, the
map $d$ is determined by $(\beta_{1,1},\beta_{1,2},\beta_{1,3})$.
There exist a symmetric analog of Proposition~\ref{17} rephrasing the
condition~(\ref{6}) of Definition~\ref{16} in
terms of $(\beta_{1,1},\beta_{1,2},\beta_{1,3})$,
but it would make the paper too long so we prefer to proceed
immediately to examples.

\begin{example}{\rm\
\label{46}
In this example we give an innocuous generalization
of such classical objects as Poisson or Gerstenhaber algebras.
Let us fix two natural numbers, $m$ and $n$. Let $U$ be the
graded vector space spanned on an element $\mu$ of degree $m$ and
let $V$ be the graded vector space spanned on an element $\nu$ of
degree $n$. Define $\Sigma_2$-actions on $U$ and $V$ by $\sigma\mu :=
(-1)^m\cdot \mu$ and $\sigma\nu:= -(-1)^n\cdot \nu$. Let
$S\subset \fr(U)(3)$ be the $\Sigma_3$-invariant subset generated by
$\mu\c_1\mu-(-1)^m\cdot \mu\c_2\mu$ (the associativity)
and let $T \subset \fr(V)(3)$ be
the $\Sigma_3$-invariant subset generated by $\nu\c_2\nu +
(\nu\c_1\nu)(1\otimes \sigma)+(-1)^n \cdot \nu\c_1\nu$
(the Jacobi identity).
Finally, let $d: V\b U\to U\b V$ be given by $d(\nu\c_2\mu):=
\mu\c_1\nu +(-1)^m\cdot
(\mu\c_1\nu)(1\otimes \sigma)$. The reader will easily
verify that this gives a distributive law.

An algebra over the
operad $\P(m,n):= \langle U,V; S,D,T\rangle$ defined above consists of
a
(graded) vector space
$P$ together with two bilinear maps, $-\cup-:P \ot P \to
P$ of degree $m$, and $[-,-]:P\ot P\to P$ of degree $n$
such that, for any homogeneous $a,b,c\in P$,
\begin{enumerate}
\item[(i)]
$a\cup b = (-1)^{|a|\cdot |b|+m}\cdot b\cup a$,
\item[(ii)]
$[a,b] = -(-1)^{|a|\cdot |b|+n}\cdot [b,a]$,
\item[(iii)]
$-\cup-$ is associative in the sense that
\[
a\cup (b\cup c) = (-1)^{m\cdot(|a|+1)}\cdot(a\cup b)\cup c,
\]
\item[(iv)]
$[-,-]$ satisfies the following form of the Jacobi identity:
\[
(-1)^{|a|\cdot(|c|+n)}\cdot [a,[b,c]]+(-1)^{|b|\cdot(|a|+n)}\cdot
[b,[c,a]]
+(-1)^{|c|\cdot(|b|+n)}\cdot [c,[a,b]] = 0,
\]
\item[(v)]
the operations $-\cup-$ and $[-,-]$ are compatible in the sense that
\[
(-1)^{m\cdot|a|}\cdot [a,b\cup c] =
[a,b]\cup c + (-1)^{(|b|\cdot |c|+m)}\cdot [a,c]\cup b.
\]
\end{enumerate}
Following~\cite{fox-markl:preprint} we call algebras as above
$(m,n)$-{\em algebras\/}.
Obviously $(0,0)$-algebras are exactly (graded) Poisson algebras,
$(0,-1)$-algebras are Gerstenhaber algebras introduced in~\cite{ADT88}
while $(0,n-1)$-algebras
are the $n$-algebras of~\cite{getzler-jones:preprint}.
We may think of an $(m,n)$-structure on
$P$ as of a Lie algebra structure on the $n$-fold
suspension $\uparrow^n \hskip-1mm P$
of the graded vector space $P$
together with an associative commutative algebra
structure on the $m$-fold suspension $\uparrow^m \hskip-1mm P$
such that both structures are related via
the compatibility axiom~(v). For a more detailed analysis of this
example from an operadic point of view, see~\cite{fox-markl:preprint}.
}\end{example}

As an example of the application of the coherence theorem
(Theorem~\ref{13}) we give the following proposition which is
probably well-known and certainly frequently used, but we have not
seen a proof in the literature.

\begin{proposition}
Let $V$ be a graded vector space. Let us
take the free graded Lie algebra
${\bf L}(V)$ on $V$, forget the Lie bracket and then take the free
graded commutative algebra ${\mbox{\large $\land$}}({\bf L}(V))$ on
${\bf L}(V)$.

On the other hand, let ${\bf P}(V)$ be the free graded
Poisson algebra on
$V$ and let us again forget the Lie bracket on
${\bf P}(V)$. Then there
exists a natural isomorphism
\[
{\bf P}(V) \cong {\mbox{\large $\land$}}({\bf L}(V))
\]
of graded commutative associative algebras.
\end{proposition}

There is an obvious immediate generalization of the above proposition
to $(m,n)$-algebras as well as to other algebras with a distributive
law.

\section{Distributive laws and the Koszulness}

Let us discuss the associative algebra case first.
Let $A = \langle X;R\rangle$ be a
quadratic associative algebra. Let us denote by $\# X$ the $K$-linear
dual of $X$ and let $R^{\perp}\subset \#X\ot \#X \cong \#(X\ot X)$ be
the annihilator
of $R\subset X\ot X$. Then define~\cite{manin:AIF87} the
{\em Koszul dual\/} of $A$ to be the quadratic algebra $A^! :=
\langle \#X, R^{\perp}\rangle$. We have the following lemma.

\begin{lemma}
\label{18}
Let $R_{s,n}:= T^s(X)\ot R\ot T^{n-s-2}\subset F_n(X)$,
$0\leq s\leq n-2$, $n\leq 2$, be as in~\ref{36}. Then
\[
\# A^!_n = \left\{
\begin{array}{ll}
\bigcap\{R_{s,n};\ 0\leq s\leq n-2\},& \mbox{ for } n\geq 2,
\\
X, & \mbox{ for } n=1, \mbox{ and}
\\
K, & \mbox{ for $n=0$}.
\end{array}
\right.%}
\]
\end{lemma}

\noindent
{\bf Proof.}
Applying the considerations of~\ref{36} to the quadratic algebra
$A^!$ we get
\[
A_n^!= T^n(\#X)/\sp((R^{\perp})_{s,n};\
0\leq s\leq n-2),
\]
the rest is an easy linear algebra.
\qed

\vskip-8mm
\hphantom{p}
\begin{odstavec}{\rm{\bf --}
\label{42}
Let us recall the definition of the {\em Koszul complex\/} $K_\b(A) =
(K_\b(A), d_A)$~\cite{manin:AIF87}.
It is a chain complex with $K_n(A):= A\ot \#A_n^!$ and the
differential $d_A$ defined as follows. For $a\ot x_1\ot \cdots\ot x_n
\in A \ot T^n(X)$ put $d(a\ot x_1\ot \cdots\ot x_n):=
a\cdot [x_1]\ot x_2\cdots \ot x_n$, where $[x_1]$ is the image of
$x_1\in X$ under the composition $X \subset T(X) \to A= T(X)/(R)$. By
Lemma~\ref{18}, $K_\b(A)\subset A\ot T(X)$ and we define $d_A$ to be
the restriction of $d$ to $K_\b(A)$. We say that $A$ is
{\em Koszul\/}~\cite{manin:AIF87} if $H_n(K_\b(A))=0$ for $n\geq 1$
and
$H_0(K_\b(A))=\bk$.
\eodst}\end{odstavec}

Let us discuss the Koszulness for associative algebras with a
distributive law. The reader will easily prove the following lemma
which we need in the sequel.

\begin{lemma}
\label{22}
Suppose $C = \langle U,V; S,D,T\rangle$ is a quadratic
algebra with a distributive
law $d:V\ot U\to U\ot V$. Let $\#d :\#U\ot \#V\to \#V \ot \#U$ be the
dual of $d$ and let $D^\perp := \{\alpha \ot \beta -
\#d(\alpha\ot \beta);\ \alpha \ot \beta \in \#U\ot \#V\}$. Then $C^!
= \langle \#V,\#U; T^\perp,D^\perp, S^\perp\rangle$ and $\#d$ is a
distributive law.
\end{lemma}

Let $K_\b(C)$ be the Koszul complex of an algebra
$C = \langle U,V; S,D,T\rangle$ with a distributive law. As $C^!
= \langle \#V,\#U; T^\perp,D^\perp, S^\perp\rangle$, we have the
bigrading $C^!_n = \bigoplus_{i+j=n}C^!_{i,j}$ with $C^!_{1,0}=\#V$,
$C^!_{0,1}=\#U$, which induces the
decomposition $\#C^!_n = \bigoplus_{i+j=n}\#C^!_{i,j}$ with
$\#C^!_{1,0}=V$ and $\#C^!_{0,1}=U$. We may
define the convergent decreasing filtration $F_\b K_\b(C)$ of
$K_\b(C)$ by
$F_pK_{n}(C):= \bigoplus _{i\leq p}C\ot \#C^!_{n-i,i}$.
Obviously $d_C(F_p K_n(C))\subset F_p K_{n-1}(C)$, hence the
filtration
induces a first quadrant
spectral sequence $\E = \{E^r_{p,q},d^r\}$ which converges
to $H_\b(K_\b(C))$.

\begin{proposition}
\label{19}
For the spectral sequence $\E=\{E^r_{p,q},d^r\}$ above
we have the following
isomorphism of differential graded modules:
\[
(E^0_{p,\b},d^0) \cong (A\ot K_\b(B)\ot \#A^!_p, \id \ot d_B
\ot \id).
\]
\end{proposition}

\noindent
{\bf Proof.}
We have, by Theorem~\ref{8}, a natural isomorphism
$\xi : A\ot B\to C$. Because
$C^!$ is, by Lemma~\ref{22}, also an algebra with a distributive law,
we
have the isomorphism $\xi_{q,p}:B^!_q \ot A^!_p
\to C^!_{q,p}$ which induces the dual isomorphism $\#\xi_{q,p}:\#
C^!_{q,p} \to\#B^!_q \ot \#A^!_p $. On the other hand, we have an
obvious identification $E^0_{p,q} = C\ot \#C^!_{q,p}$ and we
define, using this identification, the isomorphism
$\phi_{p,q}:E^0_{p,q}
\to A\ot B \ot \#B_q^! \ot \#A^!_p $ by $\phi_{p,q}:=
\xi^{-1}\ot \#\xi_{q,p}$. We must show that this map commutes with
the differentials, i.e. that for $z \in C\ot \#C^!_{p,q}$,
\begin{equation}
\label{20}
\phi_{p,q-1}(d^0(z))= (\id\ot d_B\ot \id)(\phi_{p,q}(z)).
\end{equation}

Let us observe first that there is a very explicit way to
describe the map
$\#\xi_{q,p}$ using the identification of $\#C^!_{q,p}$ with
a subspace of $F_{q,p}(V,U)$. If $\pi :
F_{q,p}(V,U)\to T^q(V)\ot T^p(U)$ is the projection, then
$\#\xi_{q,p}$ coincides with the restriction of $\pi$ to
$\#C^!_{q,p}$.

Suppose now that, in~(\ref{20}), $z = x\ot y$, $x\in C$ and
$y\in \#C^!_{q,p}$. We may then write $x = a\cdot b$ for some
$a\in A$ and $b\in B$, and $y= w+ \sum v_\omega \ot y_\omega$ with
$y_\omega \in F_{q-1,p}$, $v_\omega \in V$, and $w
\in U\ot F_{q,p-1}(V,U)$. We then have
$\phi_{p,q-1}(d^0(x\ot y))= \phi_{p,q-1}(\sum a\cdot b
\cdot [v_\omega]\ot y_\omega)= \sum \xi^{-1}(a\cdot b
\cdot [v_\omega])\ot \#\xi_{p,q-1}(y_\omega) = \sum a\ot (b
\cdot [v_\omega])\ot \#\xi_{p,q-1}(y_\omega)$ , while $(\id \ot d_B
\ot\id)(\phi_{p,q}(x\ot y))= \sum (\id \ot d_B
\ot\id)(\xi^{-1}(a\cdot b)\ot
v_\omega \ot \#\xi_{p,q-1}(y_\omega)) = \sum a\ot (b
\cdot [v_\omega])\ot \#\xi_{p,q-1}(y_\omega)$ and~(\ref{20}) follows.
We used the obvious equality $\xi^{-1}(c\cdot d)= c\ot d$ for any
$c\in A$ and $d\in B$.
\qed

As a corollary we obtain the following theorem.

\begin{theorem}
\label{21}
Let $C = \langle U,V; S,D,T\rangle$ be an associative
algebra with a distributive law,
$A := \langle U,S\rangle $ and $B := \langle V; T\rangle$. If the
algebras $A$ and $B$ are Koszul, then $C$ is a Koszul algebra as well.
\end{theorem}

\noindent
{\bf Proof.}
The Koszulness of $B$ means that $H_q(K_\b(B), d_B)= 0$ for $q\geq 1$
and $H_0(K_\b(B), d_B)= \bk$. The K\"unneth
formula together with
Proposition~\ref{19} gives that $E_{p,q}^1= 0$ for $q\geq 1$ and that
$E_{p,0}^1= K_p(A)$. We can easily identify the differential
$d^1_{p,0}$ with $d_A$ which finishes the proof.
\qed

In the rest of this paragraph we formulate and prove an analog
of Theorem~\ref{21}
for operads. We need some notation.
For a nonsymmetric collection $\{C(n); n\geq 1\}$ we define
the {\em dual\/} $\#C$ by $(\#C)(n):= \#(C(n))$. In the symmetric
case the definition is the same with the action of $\Sigma_n$ on
$\#C(n)$
being the induced action multiplied by the sign representation. In
both
cases we have a canonical isomorphism of collections $\#\fr(C)=
\fr(\#C)$. The {\em Koszul dual\/} $\S^!$ of a quadratic operad $\S =
\langle E;R\rangle$ is then,
following~\cite{ginzburg-kapranov:preprint}
defined as $\S^!:= \langle \#E;R^\perp\rangle$, where
$R^\perp\subset \fr(\#E)(3)= \#\fr(E)(3)$ is the annihilator of
the subspace $R\subset \fr(E)(3)$. We have the following analog of
Lemma~\ref{18}.

\begin{lemma}
\label{23}
Let $R_{\tS}\subset \fr(E)$ be, for $\tS\in \trb_n$,
the same as in~\ref{37}. Then
\[
\# \S^!(n) = \left\{
\begin{array}{ll}
\bigcap\{R_{\tS};\ \tS\in \trb_n\},& \mbox{ for } n\geq 3,
\\
E, & \mbox{ for } n=2, \mbox{ and}
\\
\bk, & \mbox{ for $n=1$}.
\end{array}
\right.%}
\]
\end{lemma}

\noindent
{\bf Proof.}
The same linear algebra as in the proof of Lemma~\ref{18}.
For the symmetric case the
statement was formulated in~\cite{ginzburg-kapranov:preprint}.
\qed

\begin{odstavec}{\rm{\bf --}
\label{43}
We are going to define the Koszul complex of an operad, rephrasing,
in fact, a definition of~\cite{ginzburg-kapranov:preprint}.
The {\em Koszul complex\/} of an operad $\S=\prez ER$
is a differential
graded collection $K_\b(\S)= (K_\b(\S),d_\S)$ with $K_\b(\S):=
\S\cd \#\S^!$. The component $K_n(\S)(m)
\subset (\S \cd \#\S^!)(m)$ is generated by elements of the form
$\gamma(s;t_1,\ldots,t_k)$, $s\in \S(k)$, $t_i \in \#\S^!_{j_i}(m_i)$,
$1\leq i\leq k$, where $m_1+\cdots+m_k = m$ and $j_1+\cdots+j_k=n$.
As $\#\S^! \subset \fr(U)$ by Lemma~\ref{23}, we may
in fact suppose that $t_i \in \fr(E)$ (or, in a more compact
notation, that $K_\b(\S)\subset \S\cd \fr(E)$).
The differential is defined as follows.
Let $x = \gamma(s;t_1,\ldots,t_k)$, $t_i\in \fr(E)(m_i)$
be as above. If $m_i=1$ put $d_i(x)=0$. For $m_i> 1$, $x$ can be
obviously rewritten as $x=\gamma(s\c_ir_i;y_1,\ldots,y_{k+1})$ with
$r_i\in E$ and with some $y_1,\ldots,y_{k+1}\in \fr(E)$ (in fact,
$y_j=t_j$ for $j< i$ and $y_{j+1}= t_j$ for $j>i$). Define then
$d_i:= \gamma(s\c_i[r_i]; y_1,\ldots,y_{k+1})$,
where $[-]:E \to \S$ maps $e\in E$ to its class $[e]$ in $\S=
\fr(E)/(R)$.
Then we put $d(x):= \sum_{1\leq i\leq k}d_i(x)$.
The differential $d_{\S}$ on $K_\b(\S)$ is defined as the restriction
of $d$
to $K_\b(\S)\subset \S\c \fr(E)$. We can verify that
$d_{\S}^2=0$; for the symmetric case it was done
in~\cite{ginzburg-kapranov:preprint}, the
nonsymmetric case is even easier.
As in~\cite{ginzburg-kapranov:preprint} we say that $\S$ is
{\em Koszul\/} if the complex $(K_\b(\S)(m),d_{\S}(m))$ is acyclic
for any $m\geq 2$. Observe that, by definition, $K_\b(\S)(1)=
K_0(\S)(1)=\bk$.
\eodst}\end{odstavec}

Before discussing the Koszulness of operads with a distributive law
we state the following analog of Lemma~\ref{22} which was formulated
in~\cite{fox-markl:preprint}, the verification is immediate.

\begin{lemma}
\label{24}
Let $\C = \langle U,V; S,D,T\rangle$ be an operad with a distributive
law $d:V\b U\to U\b V$. Let $\#d :\#U\b \#V\to \#V \b \#U$ be the
dual of $d$ and let $D^\perp := \{\alpha \b \beta -
\#d(\alpha\b \beta);\ \alpha \b \beta \in \#U\b \#V\}$. Then $\C^!
= \langle \#V,\#U; T^\perp,D^\perp, S^\perp\rangle$ and $\#d$ is a
distributive law.
\end{lemma}

Consider the Koszul complex $K_\b(\C)$ of an operad $\C =
\langle U,V;S,D,T\rangle$ with a distributive law. By Lemma~\ref{24},
$\C^! = \langle \#V,\#U; T^\perp,D^\perp, S^\perp\rangle$, and we
have the bigrading $\C^!_n = \bigoplus_{i+j=n} \C^!_{i,j}$ which
induces the decomposition $\#\C^!_n = \bigoplus_{i+j=n} \#\C^!_{i,j}$
with $\#\C^!_{1,0}=V$ and $\#\C^!_{0,1}=U$
of the dual collection. We may use these data to define the
convergent decreasing filtration $F_\b K_\b(\C)$ of $K_\b(\C)$ as
follows. Let $F_pK_n(\C)\subset K_n(\C)$
be generated by elements
$\gamma(s;t_1,\ldots,t_k)$, $s\in \C(k)$, $t_i\in \#\C^!_{a_i,b_i}$,
$1\leq i\leq k$, $\sum_{i=1}^kb_i \leq p$ and
$\sum_{i=1}^k(a_i+b_i)=n$. We can easily see that the differential
$d_\C$ preserves the filtration, $d_\C(m)F_pK_n(\C)(m)
\subset F_pK_{n-1}(\C)(m)$, therefore there is a spectral sequence
${\bf E}(m) = (E^r_{p,q}(m), d^r(m))$ converging to
$H_\b(K_\b(\C))(m)$, for any $m\geq 1$.

Let us observe that, for any three collections $X,Y$ and $Z$,
the collection $X\cd Y\cd Z$ is naturally bigraded by
$(X\cd Y\cd Z)_{p,q}:= \fr_{1,p,q}(X\op Y\op Z)\cap (X\op Y\op Z)$
and we write $X\cd Y_p\cd Z_q$
instead of $(X\cd Y\cd Z)_{p,q}$. We have the following analog of
Proposition~\ref{19}.

\begin{proposition}
For the spectral sequence ${\bf E}(m) = (E^r_{p,q}(m), d^r(m))$
defined above we have, for each $m\geq 1$,
\[
(E^0_{p,\b}(m), d^0(m)) \cong ((\A\cd K_\b(\B)\cd \#\A^!_p)(m),
(\id \cd d_\B \cd \id)(m)).
\]
\end{proposition}

\noindent
{\bf Proof.}
By Theorem~\ref{13} we have the isomorphism of collections $\xi:
\A\cd B\to \C$ and, because $\C^!$ is, by Lemma~\ref{24}, also an
operad with a distributive law, by the same theorem we have an
isomorphism $\xi :\B^!\cd \A^! \to \C^!$ inducing the dual
isomorphism $\#\xi_{q,p}: \#\C^!_{q,p}\to \#\B^!_q\cd \#\A^!_p$ of
bigraded collections.

We have the identification $E^0_{p,q}=\C\cd \C^!_{q,p}$ (=
the space generated by elements $\gamma(s;t_1,\ldots,t_k)$ with
$s\in \C(k)$, $t_i\in \#\C^!_{a_i,b_i}$, $1\leq i\leq k$,
$\sum_{i=1}^k a_i = q$ and $\sum_{i=1}^k
b_i =p$). We may thus define an isomorphism of collections
$\phi_{p,q}:E^0_{p,q}\to \A\cd \B\cd \#\B^!_q\cd \#\A^!_p$ by
$\phi_{p,q}:= \xi^{-1}\cd \#\xi_{q,p}$. We must show that this map
commutes with the differential, i.e.~that, for
$z\in \C\cd \#\C^!_{p,q}$,
\begin{equation}
\label{41}
\phi_{p,q-1}(d^0(z))= (\id\cd d_\B\cd \id)(\phi_{p,q}(z)).
\end{equation}
We have, similarly as in the associative algebra case, a very
explicit description of $\#\xi_{q,p}$ given as follows.
As in~\ref{37}, $\fr(V,U)= \bigoplus_{\tT \in \tr^{wb,2}}(V,U)(\tT)$
which gives the canonical direct sum decomposition $\fr(V,U)=
\bigoplus_{l\geq 0}\fr(V,U)_{(l)}$ with $\fr(V,U)_{(l)}:=
\bigoplus_{\tT \in \tr^{wb,2},I(\tT)=l}(V,U)(\tT)$. Observing that
$\fr(V,U)_{(0)}= \fr(V)\cd \fr(U)$ we conclude that
$\fr_q(V)\cd \fr_p(U)$ is a canonical direct summand of
$\fr_{q,p}(U,V)$. Let $\pi : \fr_{q,p}(V,U)\to \fr_q(V)\cd \fr_p(U)$
be the corresponding projection.
Using the identification of $\#\C^!_{q,p}$ with a subspace of
$\fr_{q,p}(V,U)$ provided by Lemma~\ref{23},
the map $\#\xi_{q,p}$ coincides with the
restriction of $\pi$ to $\#\C^!_{q,p}$.

The remaining arguments are similar to those in the proof of
Proposition~\ref{21} but much more technically complicated. It is
obviously enough to prove~(\ref{41}) for elements $z$ of the form $z=
\gamma(s;t_1,\ldots,t_k)$ with $t_i\in \#\C^!_{a_i,b_i}$ for
$1\leq i\leq k$,
$\sum_{i=1}^k a_i = q$ and $\sum_{i=1}^k b_i =p$. We may also suppose
that $s = \gamma(a;b_1,\ldots,b_l)$ for some $a\in \A(l)$,
$b_i\in \B(m_i)$, for $1\leq i\leq l$ and $\sum_{i=1}^l m_i=k$. We
may also take $t_i$ to be of the form $t_i = w_i + r_i$ with $r_i =
\sum_{\omega \in \Omega_i}\gamma(v_{i,\omega};
y_{i,\omega,1},y_{i,\omega,2})$, for $v_{i,\omega}\in V$ and
$y_{i,\omega,j}\in \#\C^!_{a_{i,\omega,j},b_{i,\omega,j}}$, $j=1,2$,
with $a_{i,\omega,1}+a_{i,\omega,2}= a_i-1$ and $b_{i,\omega,1}+
b_{i,\omega,2}= b_i$, and $w_i \in U\cd \fr_{a_i,b_i-1}(V,U)$. Any
other element $z\in\C\cd \#\C^!_{p,q}$ can be expressed as a linear
combination (and using the symmetric group action in the symmetric
case) of elements of the above form.

We also denote, for each $1\leq i\leq k$, by $s(i)$ the unique number
such that $m_1+\cdots+m_{s(i)-1}< i\leq m_1+\cdots+m_{s(i)}$, let us
then put $t(i):= i- m_1+\cdots+m_{s(i)-1}$. For any given $i$,
$1\leq i \leq k$, we have
\begin{eqnarray*}
z&=& \sum_{\omega\in \Omega_i}
\alpha\cdot \gamma(s\c_i
v_{i,\omega};t_1,\ldots,t_{i-1},y_{i,\omega,1},y_{i,\omega,2},
t_{i+1},\ldots,t_k)
\\
&&+\gamma(s;t_1,\ldots,t_{i-1},w_i,t_{i+1},\ldots,t_k)
\end{eqnarray*}
therefore
\begin{eqnarray*}
\phi_{p,q-1}(d^0(z)) &=&\sum_{\omega\in\Omega_i,1\leq i\leq k}
\alpha\cdot
\phi_{p,q-1}(\gamma(s\c_i
[v_{i,\omega}];t_1,\ldots,t_{i-1},y_{i,\omega,1},y_{i,\omega,2},
t_{i+1},\ldots,t_k))
\\
&=&\sum_{\omega\in\Omega_i,1\leq i\leq k}
\alpha\beta\cdot\gamma(\gamma(a;b_1,\ldots,b_{s(i)-1},b_{s(i)}
\c_{t(i)}[v_{i,\omega}],
b_{s(i)+1},\ldots,b_l);
\\
&&\hskip2.3cm\#\xi(r_1),\ldots,\#\xi(r_{i-1}),
\#\xi(y_{i,\omega,1}),\#\xi(y_{i,\omega,2}), \#\xi(r_{i+1}),\ldots,
\#\xi(r_k))
\end{eqnarray*}
with $\alpha:= (-1)^{|t_1|+\cdots+|t_{i-1}|}$ and $\beta:=
(-1)^{|b_{s(i)+1}|+\cdots+|b_l|}$.
Here we used the clear fact that $\#\xi(t_i)= \#\xi(r_i)$.
On the other hand,
\[
(\id\cd d_\B\cd\id)(\phi_{p,q}(z))= (\id\cd d_\B\cd\id)
\gamma(\gamma(a;b_1,\ldots,b_l);\#\xi(r_1),\ldots,\#\xi(r_k))
\]
and this expression coincides, taking
into the account the relation
\[
\#\xi(r_i)=
\sum_{\omega \in \Omega_i}\#\xi(\gamma(v_{i,\omega};
y_{i,\omega,1},y_{i,\omega,2}))= \sum_{\omega
\in \Omega_i}\gamma(v_{i,\omega};
\#\xi (y_{i,\omega,1}),\# \xi( y_{i,\omega,2})),
\]
with the right-hand side term of the equation above.
\qed

The following theorem which is, in fact, one of the central results
of the paper, easily follows from the previous proposition and from
the K\"unneth formula for collections (Proposition~\ref{38}).

\begin{theorem}
\label{44}
Let $\C = \langle U,V; S,D,T\rangle$ be an operad with a distributive
law and let $\A := \langle U;S\rangle$ and $\B :=
\langle V;T\rangle$. If the operads $\A$ and $\B$ are Koszul, then
$\C$ is Koszul as well.
\end{theorem}

Before discussing some immediate consequences of Theorem~\ref{44},
let us make one more comment. For an operad $\S$, let
${\bf s}\hskip1mm\S$ (the
{\em suspension\/}) be the operad with $({\bf s}\hskip1mm\S)(n):=
\uparrow^{n-1}\S(n)$, $n\geq 1$, with the
composition maps defined in an
obvious way; here $\uparrow^{n-1}$ denotes the usual $(n-1)$-fold
suspension of a graded vector space. It follows from the computation
of~\cite{zebrulka} that $\S$ is Koszul if and only if
its suspension ${\bf s}\hskip1mm\S(n)$ is Koszul.

Let $\P(m,n) = \langle U,V; S,D,T\rangle$ be the operad for
$(m,n)$-algebras as in Example~\ref{46}. It is immediate to see that
$\A = {\bf s}^n \mbox{\it Comm}$ and that
$\B = {\bf s}^m \mbox{\it Lie}$
while both {\it Comm} (the operad for
commutative associative algebras) and
{\it Lie} (the operad for Lie algebras) are well-known to be
Koszul, see~\cite{ginzburg-kapranov:preprint}.
Theorem~\ref{13} then gives as a corollary the following statement.

\begin{corollary}
\label{50}
The operad $\P(m,n)$ for $(m,n)$-algebras is Koszul for any two
natural numbers
$m$ and $n$. Especially, the operad $\P(0,0)$ for Poisson algebras,
the operad $\P(0,-1)$ for Gerstenhaber algebras and the operad
$\P(0,n-1)$ for $n$-algebras are Koszul.
\end{corollary}

\begin{example}{\rm\
In this example
we use the notation introduced in Example~\ref{47}. The operad
$\prez E{R_{[a,b]}}$ is Koszul for
$[a,b]\in \{[0,0],[1,1],[\infty,\infty]\}$; the values $[0,0]$ and
$[\infty,\infty]$ are trivial extreme cases where the Koszulness can
be verified directly while $[1,1]$ corresponds to the nonsymmetric
operad for associative algebras which is known to be Koszul,
see~\cite{ginzburg-kapranov:preprint}. We may then conclude that the
operad characterized by a triple
$([a_U,b_U],(\aa12,\aa21),[a_V,b_V])$ is Koszul if
\begin{enumerate}
\item
at least one of the conditions (i)--(iv) of Example~\ref{47} is
satisfied, and
\item
$[a_U,b_U],[a_V,b_V]\in \{[0,0],[1,1],[\infty,\infty]\}$.
\end{enumerate}
Especially, the `nonsymmetric Poisson algebra' of Example~\ref{47} is
Koszul.
}\end{example}

\catcode`\@=11
\noindent
Address: Mathematical Institute of the Academy, \v Zitn\'a 25, 115 67
Praha 1, Czech Republic,\hfill\break\noindent
\hphantom{Address:} email: {\bf markl@earn.cvut.cz}


\begin{thebibliography}{10}

\bibitem{beck:LNM80}
J.~Beck.
\newblock Distributive laws.
\newblock {\em Lecture Notes in Mathematics}, 80:119--140, 1969.

\bibitem{fox-markl:preprint}
T.F.~Fox and M.~Markl.
\newblock Distributive laws and the cohomology.
\newblock In preparation.

\bibitem{fulton-macpherson:anm94}
W.~Fulton and R.~MacPherson.
\newblock A compactification of configuration spaces.
\newblock {\em Annals of Mathematics}, 139:183--225, 1994.

\bibitem{ADT88}
M.~Gerstenhaber and S.D. Schack.
\newblock Algebraic cohomology and deformation theory.
\newblock In {\em Deformation Theory of Algebras and Structures and
 Applications}, pages 11--264. Kluwer, Dordrecht, 1988.

\bibitem{getzler-jones:preprint}
E.~Getzler and J.D.S. Jones.
\newblock Operads, homotopy algebra,
and iterated integrals for double loop spaces.
\newblock Preprint, 1993.

\bibitem{ginzburg-kapranov:preprint}
V.~Ginzburg and M.~Kapranov.
\newblock Koszul duality for operads.
\newblock Preprint, 1993.

\bibitem{KSV}
{T. Kimura, J.D. Stasheff and A.A. Voronov.}
On operad structures on moduli spaces and string theory. Preprint RIMS
936, 1993.

\bibitem{manin:AIF87}
Y.I. Manin.
\newblock Some remarks on {Koszul} algebras and quantum groups.
\newblock {\em Annales de l'Institut Fourier}, 37.4:191--205, 1987.

\bibitem{zebrulka}
M.~Markl.
\newblock Models for operads.
\newblock Preprint, 1994.

\bibitem{may:1972}
J.P.~May.
\newblock {\em The Geometry of Iterated Loop Spaces}, volume 271 of
{\em
 Lecture Notes in Mathematics}.
\newblock Springer-Verlag, 1972.

\bibitem{smirnov:USSRSb82}
V.A.~Smirnov.
\newblock On the cochain complex of topological spaces.
\newblock {\em Math. USSR Sbornik}, 43:133--144, 1982.

\end{thebibliography}
\end{document}